\documentclass[sigconf]{acmart}

\usepackage{amsmath}
\usepackage{multirow}
\usepackage{subcaption}
\usepackage{algorithm}
\usepackage{algpseudocode}

\renewcommand\footnotetextcopyrightpermission[1]{}

\newcommand{\todo}[1]{\textcolor{red}{[#1]}}

\newcommand{\ignore}[1]{{}}

\newcommand{\yimin}[1]{#1}

\renewcommand{\vec}[1]{\mathbf{#1}} 
\newcommand{\mat}[1]{\mathbf{#1}} 

\newcommand{\blue}{\color{blue}}

\newlength{\subcolumnwidth}

\newcommand{\nextsubcolumn}[1][]{%
  \cr\noalign{\hfill}
  \if\relax\detokenize{#1}\relax\else\hsize=#1\setlength{\subcolumnwidth}{\hsize}\fi
}

\newcommand{\sect}{\textsection}

\setcopyright{acmcopyright}

\begin{document}

\title{Interpersonal Distance Tracking with mmWave Radar and IMUs}


\author{Yimin Dai}
\affiliation{%
  \department{School of Computer Science and Engineering}
  \institution{Nanyang Technological University}
  \country{Singapore}
}

\author{Xian Shuai}
\affiliation{%
  \department{Department of Information Engineering}
  \institution{The Chinese University of Hong Kong}
  \state{Hong Kong SAR}
  \country{China}
}

\author{Rui Tan}
\affiliation{%
  \department{School of Computer Science and Engineering}
  \institution{Nanyang Technological University}
  \country{Singapore}
}

\author{Guoliang Xing}
\affiliation{%
  \department{Department of Information Engineering}
  \institution{The Chinese University of Hong Kong}
  \state{Hong Kong SAR}
  \country{China}
}


\begin{abstract}
  Tracking interpersonal distances is essential for real-time social distancing management and {\em ex-post} contact tracing to prevent spreads of contagious diseases. Bluetooth neighbor discovery has been employed for such purposes in combating COVID-19, but does not provide satisfactory spatiotemporal resolutions. This paper presents ImmTrack, a system that uses a millimeter wave radar and exploits the inertial measurement data from user-carried smartphones or wearables to track interpersonal distances. By matching the movement traces reconstructed from the radar and inertial data, the pseudo identities of the inertial data can be transferred to the radar sensing results in the global coordinate system. The re-identified, radar-sensed movement trajectories are then used to track interpersonal distances. \yimin{In a broader sense, ImmTrack is the first system that fuses data from millimeter wave radar and inertial measurement units for simultaneous user tracking and re-identification.} Evaluation with up to 27 people in various indoor/outdoor environments shows ImmTrack's decimeters-seconds spatiotemporal accuracy in contact tracing, which is similar to that of the privacy-intrusive camera surveillance and significantly outperforms the Bluetooth neighbor discovery approach.
\end{abstract}




\maketitle

\section{Introduction}
\label{sec:intro}



Retrospective studies have shown that infectious control measures including wearing masks, hand hygiene, and interpersonal distancing contribute to the prevention of COVID-19 and also to the decline of influenza, enterovirus, and all-cause pneumonia \cite{chiu2020impact}. When the mask-on requirement is gradually lifted during the current stage of the COVID-19 pandemic, interpersonal distancing is important to reducing  personal health risks and societal costs in healthcare.


This paper aims to design a system for interpersonal distance tracking for moving people in relatively enclosed environments that require extra attention to airborne transmissions of pathogens via respiratory droplets.
The tracking results can be used to detect unsafe contacts and generate real-time or {\em ex-post} alerts to the engaged individuals. 
COVID-19 contact tracing often adopts a spatiotemporal definition of contact, i.e., whether a questioned person spent more than $\tau$ seconds within $x$ meters from an infectious source, where the thresholds $\tau$ and $x$ can be updated according to the evolving understanding on virus transmissions. It has been commonly accepted that risk of transmission is greatest within one meter distance. In addition, SARS-CoV-2 has been found transmissible via a fleeting encounter \cite{fleeting-transmission}. The above suggest that effective contact tracing requires decimeters-seconds spatiotemporal accuracy.



Bluetooth neighbor discovery (BND) is the prevailing solution for smartphone- or wearable-based contact tracing \cite{kindt2021reliable,traceband}. However,
as analyzed in \cite{kindt2021reliable}, BND suffers 1) poor temporal resolution due to long discovery latency and 2) inaccurate distance estimation due to multipath and attenuation effects. As such, the BND-based Google/Apple Exposure Notifications System \cite{GAEN} cannot reliably detect contacts shorter than five minutes \cite{kindt2021reliable}.
The existing indoor localization techniques are in general incompetent for contact tracing. As summarized in \cite{xiao2016survey}, device-free approaches, in which the user does not carry a device, face the {\em anonymity} problem in tracking multiple users, i.e., the approaches cannot identify individual users. Without (pseudo) identities, the tracking results cannot be used for contact tracing. On the other hand, smartphone-based approaches have respective limitations, e.g., requiring dense Bluetooth beacons, privacy-intrusive due to visual sensing \cite{xu2020edge}, and insufficient accuracy of WiFi- or geomagnetism-based localization \cite{kotaru2015spotfi,he2017geomagnetism}.


\begin{figure}
  \centering
  \includegraphics[width=\linewidth]{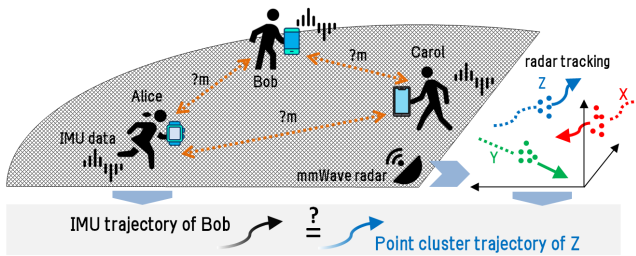}
  \vspace{-2.2em}
  \caption{ImmTrack for interpersonal distance tracking.}
  \label{fig:scenario}
  \vspace{-0.7em}
\end{figure}

Recently, millimeter wave (mmWave) radars emerged as a low-cost sensing modality and have been adopted for human detection and tracking \cite{wu2020mmtrack,shuai2021millieye}. The following features of mmWave radars form a good basis for achieving accurate interpersonal distance tracking. First, mmWave radars directly provide the velocity and depth information of the targets, which facilitate tracking the targets' absolute positions. Second, an mmWave radar can cover a large area with good sensitivity. For instance, the Texas Instruments AWR1843 mmWave radar gives a $0.23\,\text{m}$ sensing resolution within a 118\textdegree{} circular sector area with a radius of $40\,\text{m}$, covering a total area of more than 1,600$\,\text{m}^2$. Third, compared with cameras, mmWave radars output coarse-grained point clouds, which are less privacy-sensitive, making the deployment less intrusive.

However, mmWave radars also face the anonymity problem. Although research has attempted to apply supervised learning to identify the human subjects from mmWave radar data based on gaits \cite{yang2020mu}, training data from each user is needed, incurring undesirable deployment overhead. The key idea of this paper, which is illustrated in Fig.~\ref{fig:scenario}, is to exploit the inertial measurement units (IMUs) carried by the users to address the mmWave radar sensing's anonymity problem. This is based on the observations that (i) IMU data from the users inherently carry pseudo identities (PIDs), and (ii) both IMU and mmWave radar data contain rich information about the users' movements.
\yimin{While using IMU data only is insufficient for accurate tracking due to the error accumulation problem, by matching IMU data and mmWave radar data in terms of the consistency between their captured velocities and movements, the IMUs' PIDs can be transferred to mmWave radar's accurate tracking results.}
The re-identified, radar-sensed user trajectories can then be used for interpersonal distance tracking. Since IMUs are pervasively available on portable and wearable smart devices, the only requirement to enable the tracking is to share a summary of the IMU data regarding the user's movements.

Based on the above idea, we design a system called {\em {\em ImmTrack}} that employs one or more mmWave radar(s) and exploits the IMUs on the user-carried smartphones or wearables to achieve accurate interpersonal distance tracking. The design of ImmTrack needs to address the following two challenges. First, the point cloud from the radar is usually sparse and noisy \cite{shuai2021millieye}, making it difficult to separate and track multiple users during their close contacts. Second, as radar and IMU capture different aspects of movements, the cross-modality matching is non-trivial. Specifically, the radar's point cloud indicates user's space occupancy and radial movement of torso, while the IMU time series data captures linear acceleration and angular speed of the IMU-carrying limb. As a result, a common  representation of the movement inferred from the two modalities is needed for robust cross-modality matching.


To address the first challenge, ImmTrack clusters the point cloud in a single frame from the radar with initial centroids predicted by Kalman filters that capture the users' motions. The motion-aware clustering effectively prevents the wrong merge of the clusters of two users when they move close to each other. Moreover, we design a deep neural network called {\em mmClusterNet} to extract each cluster's feature capturing both shape and motion information. Then, the Hungarian algorithm associates the same user's clusters in two consecutive frames based on feature similarity, achieving multi-user inter-frame tracking. To address the second challenge, we employ a novel representation of the user's movement, called {\em trace map}, which is inferred from either the radar's tracking or the IMU's dead reckoning. We devise a Siamese neural network to extract a comparative feature from the trace map, such that the cosine similarity between two comparative features from the two modalities indicates whether they are from the same user.

We conduct experiments with up to 27 people to evaluate ImmTrack in various environments, including sports hall, lab space, and playground. Compared with the camera-based system, ImmTrack achieves similar user tracking accuracy but only incurs 1/4 to 1/2 computation overhead to process sensor data. For interpersonal distance estimation, ImmTrack achieves an average error of $22\,\text{cm}$. For pinpointing contacts within one meter over two seconds or more, ImmTrack achieves 90\% precision and 94\% recall. Compared with BND, ImmTrack reduces detection latency by up to 80 seconds. In sum, ImmTrack is suitable for hotspot venues that require extra care in preventing virus transmissions over close contacts.


The contributions of this paper are summarized as follows.
\begin{itemize}
\item We design a motion-aware mmWave radar point cloud clustering algorithm and mmClusterNet neural network for extracting cluster feature, which work together to achieve robust multi-user inter-frame tracking.
\item We propose trace map, a new modality-agnostic representation of human movement, and devise a Siamese neural network to extract feature from the trace map for effective mmWave-IMU matching.
\item \yimin{The above two designs make ImmTrack the first system that fuses data from mmWave radar and IMUs for simultaneous user tracking and re-identification.} From extensive evaluation with up to 27 people in various environments, ImmTrack achieves decimeters-seconds spatiotemporal accuracy in contact tracing.
\end{itemize}

{\em Paper organization:} \sect\ref{sec:related} presents the background and related work. \sect\ref{sec:problem} states the problem. \sect\ref{sec:global-tracking-design} and \sect\ref{subsec:association} present the designs of mmWave tracking and cross-modality matching, respectively. \sect\ref{sec:eval} presents the evaluation results.
\sect\ref{sec:conclude} concludes this paper.


\section{Background and Related Work}
\label{sec:related}

\subsection{mmWave Radar \& Comparison with Lidar}

\begin{figure}
  \centering
  \begin{minipage}[t]{.44\linewidth}
    \centering
    \vspace{0pt}
    \begin{tabular}{lll}
      \hline
       & \begin{tabular}[c]{@{}l@{}}AWR1843\\ radar\end{tabular} & \begin{tabular}[c]{@{}l@{}}A2M8\\ lidar\end{tabular} \\ \hline
      Dimension & 3D & 3D \\
      Range & 40m & 12m \\
      Resolution & 5cm & 1cm \\
      Noise & 3.2db & 15db \\ \hline
      \end{tabular}
      \vspace{0.2em}
      \captionof{table}
    {%
      Comparison between AWR1843 mmWave radar and A2M8 lidar.
      \label{tab:rlCompare}%
    }
  \end{minipage}
  \hspace{2em}
  \begin{minipage}[t]{.46\linewidth}
    \centering
    \vspace{0pt}
    \includegraphics[width=\linewidth]{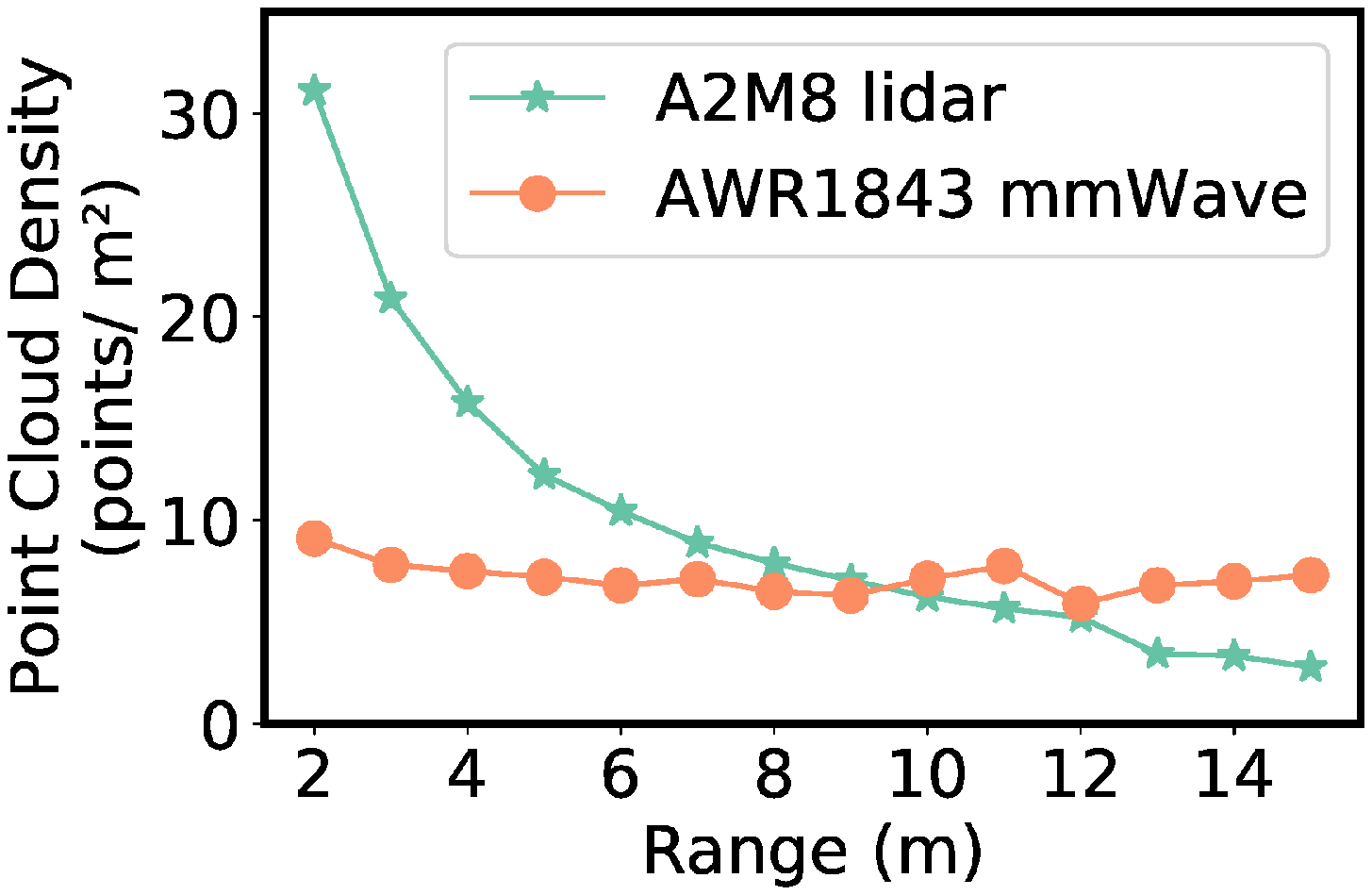}
    \vspace{-2.8em}
    \captionof{figure}
      {%
        Radar/lidar's point cloud density versus target range.
        \label{fig:denstiyCompare}%
      }%
    \end{minipage}
    \vspace{-1em}
\end{figure}


An mmWave radar can output a three-dimensional (3D) Cartesian point cloud of all the targets in the field of view (FoV), where each point is associated with a radial velocity.
Lidar and mmWave radar are often competing technologies in various applications. Lidars' higher susceptibility to occlusion, due to their short wavelengths, makes them less suitable for moving people tracking. In addition, we provide a brief comparison between the AWR1843 mmWave radar used in this paper and the A2M8 360\textdegree{} lidar used on a robot to gain more insights. The list prices of these two devices are similar. Table~\ref{tab:rlCompare} shows their sensing dimensions, ranges, resolutions, and acoustic noise levels during operation. The AWR1843 radar outperforms the A2M8 lidar except on resolution. However, the radar's $5\,\text{cm}$ resolution is satisfactory for interpersonal distance tracking.
We also measure the two devices' point cloud densities when the target's radial range varies. Fig.~\ref{fig:denstiyCompare} shows the result. As radar is mainly based on specular reflection, its point cloud density is insensitive to the target range. Differently, as lidar is mainly based on diffuse reflection, its point cloud density decreases with the target range. The attenuation may create a challenge in system design. Thus, although lidar-based interpersonal distance tracking is possible, we design ImmTrack based on mmWave radar.


\subsection{Related Work}

\subsubsection{Wireless localization, target identification, and IMU tracking}
\yimin{
Wireless indoor localization has received extensive research in the last two decades. Except camera-based surveillance, the device-free approaches in general suffer the anonymity problem in the multi-target setting \cite{xiao2016survey}.
Recently, wireless signals are used for designing device-free systems that perform both target tracking and identification. For instance, the studies \cite{zeng2016wiwho,wang2019wipin} design human subject identification systems using Wi-Fi signal. However, the used channel state information is unstable in various environments. In addition, the studies \cite{yang2020mu,zhao2019mid,pegoraro2020multiperson} train a deep learning model for human subject identification using mmWave radar signal and achieve accuracy above 92\%. However, the required extensive training data collection introduces high overheads in practice. Moreover, these systems \cite{yang2020mu,zhao2019mid,pegoraro2020multiperson} do not perform well with increased number of human subjects. This is because the distinctiveness among the radar data features is weakened when the number of human subjects increases.}
The device-based approaches, in which each target carries a signal transmitter/receiver, are free of the anonymity issue. However, as discussed in \sect\ref{sec:intro}, the device-based approaches using various modalities have respective limitations.



Besides Bluetooth, acoustic sensing is another candidate for neighbor discovery and ranging. The BeepBeep system \cite{peng2007beepbeep} performs ranging between two smartphones with audible acoustic signals.
However, the beeps in continued use is annoying.
When adapting to the near-inaudible frequency band, the operational resolution becomes unsatisfactory as the inter-device distance increases, because the smartphone audio systems are not designed to work in the inaudible band.
The studies \cite{yun2017strata} and \cite{han2016amil} that use the near-inaudible band manage to evaluate their systems when the inter-device distance is up to $0.4\,\text{m}$ and $1.2\,\text{m}$, respectively.
Thus, inaudible acoustic ranging is limited to near-field scenarios.

IMU can be used to track user's movements by dead reckoning. Embedding the resulting trajectory into the global coordinate system requires either the global coordinates of at least one point on the trajectory or certain prior knowledge like the spatial constraints expressed in the global coordinate system that any trajectory is subjected to. Dead reckoning suffers from the error accumulation problem.
A recent study \cite{ridi} applies machine learning to improve the accuracy of dead reckoning, which, however, requires massive training data and suffers domain shifts \cite{imuAdaption}.
Therefore, IMU is better for complementing other sensing modalities that can perform localization in the global coordinate system. ImmTrack uses IMUs to re-identify the mmWave radar sensing results. As ImmTrack only requires IMU's short-term dead reckoning result, it is not sensitive to the IMU dead reckoning's error accumulation problem.

\subsubsection{Multi-modality data processing}

The existing works can be classified into the following three broad categories.




\textbf{Cross-modality data translation} generates synthetic data in the target modality from real data in the source modality.
The studies \cite{videoImu,kwon2020imutube,rey2019rgbV2Imu} generate synthetic IMU data from videos of human activities. 
The work \cite{vid2doppler} generates mmWave radar data from videos. Since computer vision techniques can be employed to recognize the human activities from the videos, the synthetic IMU or mmWave radar data can be automatically labeled and used to train human activity recognition (HAR) models. 

\textbf{Multi-modal data fusion} fuses data from complementary modalities at the feature level or score/decision level to improve the robustness of sensing.
The work \cite{shuai2021millieye} fuses camera and mmWave radar to manage their respective limitations for robust object detection. Fusing camera, lidar, and radar data has been studied in the context of autonomous driving. The milliEgo system \cite{lu2020milliego} improves the accuracy of trajectory reconstruction by fusing mmWave radar data and IMU data in the single-user setting.

\textbf{Cross-modality data association} associates the sensing results in different modalities to increase information about the monitored process. The work \cite{kempfle2021quaterni} matches body-worn IMU data traces with the body joints recognized by a camera. The work \cite{ruiz2020idiot} applies the same approach to re-identify the body-worn IMUs from the video. The studies \cite{rgbw,eyefi,identitylink} associate camera data with Wi-Fi data for various purposes of augmenting the camera with depth information \cite{rgbw} or simultaneous human subject identification and tracking \cite{eyefi}. The work \cite{liuvi} associates users' smartphone Wi-Fi fine timing measurements and IMU data with a camera footage.

ImmTrack belongs to the cross-modality data association category. Different from the existing studies \cite{ruiz2020idiot,rgbw,eyefi,identitylink,liuvi} that use camera as an association source, we employ mmWave radar that is less privacy-intrusive. Moreover, technically, mmWave radar directly provides 3D locations and velocities of the human subjects, which facilitate the association.



\section{Overview of ImmTrack}
\label{sec:problem}

\subsection{Problem Description and Challenges}
\label{subsec:challenges}

We consider an enclosed space that requires extra attention to interpersonal distances, due to say the risk of airborne transmissions of pathogens via respiratory droplets. One or more mmWave radars are deployed to fully cover the space such that any human subject therein can be sensed by the radar(s). The objective of ImmTrack is to track the interpersonal distances among the users in the space. The tracking results can be sent back to the users and/or fed into downstream applications (e.g., contact tracing). When a user is about to enter the space, the user needs to enrol in ImmTrack, e.g., by quick response (QR) code scanning.
Certain user PID generation scheme can be used for ImmTrack, depending on the detailed privacy policy. For instance, the ImmTrack mobile app may generate a universally unique identifier (UUID) that takes effect throughout the lifetime of the app and is used as the PID across all ImmTrack-instrumented spaces; or the app may communicate with the ImmTrack server to generate a temporary PID that is unique in the enrolled space. The design of ImmTrack is agnostic to the PID generation scheme. When the user is in the ImmTrack-instrumented space, the ImmTrack mobile app runs in the background and collects IMU data. When the user exits the monitored space, the user needs to sign out. Thus, ImmTrack works in a nearly unobtrusive manner, except the little overhead of signing in and out incurred to the user. Such little overhead is acceptable for specific spaces that require close interpersonal distance monitoring.

The presentation of this paper focuses on a given time period, during which there are $N$ users in the monitored space. Due to the mandatory enrolment, the value of $N$, although may vary with time, is known by the system at all times.
To simplify exposition, the design presentation of ImmTrack focuses on the case that a single mmWave radar is deployed.
When a single radar is insufficient to cover the entire space, multiple radars can be deployed. The existing planning algorithms to minimize the number of cameras while achieving visual coverage \cite{he2015full,huang2013connected} can be applied to plan the radars' deployment.
\sect\ref{subsec:multi-radar} and \sect\ref{eval_tiny_space} will present the details of merging the point clouds from multiple radars and the evaluation of multi-radar ImmTrack, respectively. Although the deployment of radar(s) involves a cost, it enables the demanded close interpersonal distance monitoring. Moreover, it is a one-time cost that brings sustained benefits to the users' health and safety.

The design of ImmTrack faces the following two main challenges.

First, robust tracking of multiple users with mmWave radar is challenging. Reflections from unrelated objects may cause excessive noise points in the radar's output point cloud. Moreover, as mmWave reflections are mostly specular, the radar's point clouds are generally sparse. As such, the state-of-the-art object detection and tracking algorithms developed for processing dense point clouds yielded by high-profile lidars are ill-suited for mmWave radars.
The mmWave-based multi-user tracking also needs to deal with the users' close encounters and crossings in FoV. The DBSCAN algorithm that is widely adopted for point cloud clustering often mistakenly merges multiple users in proximity into a single cluster. As such, the clustering accuracy decreases drastically with the number of people (45\% \cite{livshitz2017tracking} and 65\% \cite{huang2021indoor} for five people). To address this issue, the clustering algorithm should maintain and incorporate the understanding of all users' movements.

Second, robust cross-modality association of the mmWave and IMU tracking results is non-trivial. The two modalities differ in the following two aspects. First, their sensing results are in different coordinate systems. Second, they capture different aspects of the user's movement. The mmWave radar captures the user torso location and velocity with lower frame rates, while the IMU captures the acceleration and angular speed of the user limb with higher frame rates. To achieve robust association, a common feature of the user's movement needs to be derived from both the mmWave data and IMU data. Moreover, the association algorithm needs to accommodate each modality's error in deriving the common feature.

\subsection{System Overview}

\begin{figure}
    \centering
    \includegraphics[width=\linewidth]{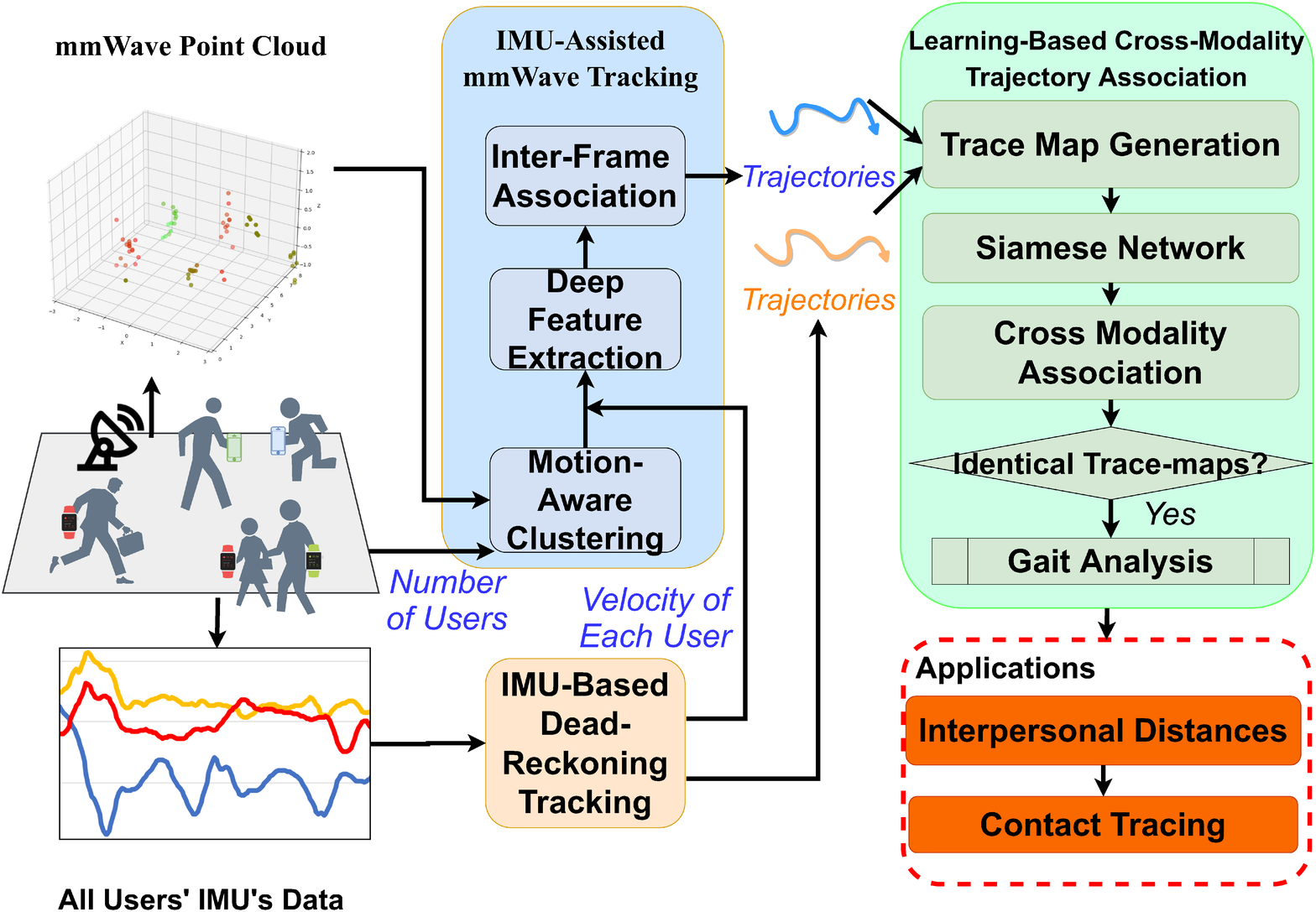}
    \vspace{-2em}
  \caption{Overview of ImmTrack. It processes data from one or more mmWave radars and users' IMUs with two components: {\em IMU-assisted mmWave tracking} and {\em learning-based cross-modality trajectory association}. The association results are fed to downstream applications.}
    \label{fig:structure}
    \vspace{-1em}
\end{figure}

Fig.~\ref{fig:structure} overviews the design of ImmTrack. It consists of two components to address the above two challenges.

{\bf IMU-assisted mmWave tracking:} This component consists of three steps. First, it clusters the points in each frame's point cloud into human bodies and associates the clusters corresponding to the same user across frames. ImmTrack maintains a recurive Kalman filter \cite{feng2014kalman} to track each user's movement and uses its predicted user location as the initial centroid of the user's cluster for the clustering algorithm. This motion-aware clustering remains robust when the users encounter each other. \yimin{Compared with the simplistic mmWave-based target tracking techniques such as that included in the radar vendor's application note \cite{livshitz2017tracking}, our algorithm avoids using heuristic object detectors such as constant false alarm rate (CFAR) detector, which easily result in detection errors.} Second, to perform the cross-frame cluster association for user tracking, a deep neural network called mmClusterNet extracts the feature of each cluster produced by the clustering algorithm. The feature incorporates the shape and motion information of the point cluster, as well as the PID of a pre-matched IMU in terms of movement velocity. \yimin{Such multidimensional information improves the robustness of the cross-frame cluster association.}
Third, the Hungarian algorithm associates the clusters across frames in terms of their features extracted by the mmClusterNet to achieve multi-user tracking. The details are presented in \sect\ref{sec:global-tracking-design}.

{\bf Learning-based cross-modality trajectory association:} ImmTrack adopts the trajectory incorporated with velocity information as the common feature of the user's movement sensed by mmWave radar and IMU. Reasons are two-fold. First, velocity-incorporated trajectory is high-level information that summarizes the user movement and generally remains consistent between the two modalities. Second, trajectory includes both temporal and spatial information. With the temporal continuity embedded in adjacent frames, the noise flickering in single frame can be largely suppressed. After the users' trajectories are reconstructed from the mmWave and IMU tracking, ImmTrack computes an imagery representation of each trajectory, which is called {\em trace map}. Then, ImmTrack applies a Siamese neural network \cite{facenet} with convolutional layers to extract comparative features from the trace maps, which are insensitive to the relative relationship between the radar's global coordinate system and the IMU's local coordinate system.
Finally, a bipartite graph matching algorithm associates the mmWave and IMU tracking results in terms of the cosine similarity among the comparative features. For users with nearly identical trace maps due to say side-by-side walks or simple straight walks, gait analysis will be performed on the involved mmWave clusters and IMU traces to generate gait features for mmWave-IMU association. The details are presented in \sect\ref{subsec:association}.

Note that except the IMU trace map generation running on each user's smartphone, all other processing tasks of ImmTrack run on an edge server or a cloud server. The smartphone transmits the periodically generated trace maps to the ImmTrack server.

\section{IMU-Assisted mmWave Tracking}
\label{sec:global-tracking-design}

\subsection{Motion-Aware Intra-Frame Clustering}
\label{subsec:mmWave-track-cluster}

\subsubsection{Design}
The radar yields a point cloud per frame. For each frame, ImmTrack removes the static points that normally correspond to the background. 
Specifically, ImmTrack compares each point's velocity with an adaptive velocity threshold updated by the triangle histogram algorithm \cite{li1998iterative} to decide whether the point is static.
ImmTrack adopts the $k$-means algorithm to divide the point cloud into $N$ clusters by setting $k=N$. 
Notably, the initial centroids often affect the performance of $k$-means.
ImmTrack uses the recursive Kalman filters (RKFs) \cite{feng2014kalman} to predict the initial centroids.

ImmTrack maintains an RKF for each user's volumetric centroid. The human body's kinetic model used by RKF is as follows. Let $\vec{x}_{i,j}$ denote the state of the $i^\text{th}$ user's centroid in the $j^\text{th}$ frame, where $i \in [1, N]$ is the internal PID in the domain of RKF. Note that this PID is different from the PID of the IMU.
We define $\vec{x}_{i,j}=[r_{i,j}, \dot{r}_{i,j}, \theta_{i,j}, \dot{\theta}_{i,j}, \phi_{i,j}, \dot{\phi}_{i,j}]^\top$, where $r_{i,j}$, $\theta_{i,j}$, and $\phi_{i,j}$ are the radial range, azimuthal and polar angles, and the overhead dot denotes the velocity. Denote by $\vec{c}_{i,j} = [\hat{r}_{i,j}, \hat{\theta}_{i,j}, \hat{\phi}_{i,j}]^\top$ the $i^\text{th}$ user's observed centroid, where the $k$-means algorithm fed with the point cloud is viewed as the observation process. By assuming that the user's velocity is constant in a frame duration (denoted by $\Delta t$), the state transition and observation models are
\begin{equation}
  \vec{x}_{i,j} = \mat{F} \vec{x}_{i,j-1} + \vec{w}_{i,j}, \quad \vec{c}_{i,j} = \mat{H} \vec{x}_{i,j} + \vec{z}_{i,j},
  \label{eq:rkf}
\end{equation}
where $\mat{F}$ is the state-transition matrix capturing the movement kinetics, $\vec{w}_{i,j}$ is the stationary process noise capturing the uncertainty of the movement, $\mat{H}$ is the observation matrix, and $\vec{z}_{i,j}$ is the non-stationary observation noise capturing the uncertainties caused by the radar's sensing noises and inaccuracy of the $k$-means algorithm. Specifically, $\mat{F} = \mathrm{diag}(\mat{A}, \mat{A}, \mat{A}) \in \mathbb{R}^{6\times 6}$, where $\mat{A} = [1, \Delta t; 0, 1]$ and $\mat{H}$ is a binary matrix that selects $r_{i,j}$, $\theta_{i,j}$, and $\phi_{i,j}$ from $\vec{x}_{i,j}$.

Before processing the $j^\text{th}$ frame, ImmTrack uses the RKF to predict the $i^\text{th}$ user's centroid $\widetilde{\vec{c}}_{i,j}$ by $\widetilde{\vec{c}}_{i,j} = \mat{H} \mat{F} \vec{x}_{i,j-1}$, where $\vec{x}_{i,j-1}$ was obtained in the previous frame. When RKF is bootstrapped (i.e., $j=0$), ImmTrack uses the DBSCAN algorithm to obtain $\widetilde{\vec{c}}_{i,0}$. Then, ImmTrack uses $\{\widetilde{\vec{c}}_{i,j} | i \in [1, N]\}$ as the initial centroids for the $k$-means algorithm with $k=N$ to process the point cloud in the $j^\text{th}$ frame. We sequentially assign the PID of each initial centroid to the closest centroid of a cluster exclusively, forming the pseudo-identified clustering result $\{\vec{c}_{i,j} | i \in [1, N]\}$. Finally, ImmTrack uses a policy derived in \cite{feng2014kalman} to update $\vec{x}_{i,j}$ and the covariance matrix of $\vec{z}_{i,j}$, i.e.,  $\vec{x}_{i,j} = \vec{x}_{i,j-1} + \vec{K}_{i,j}\left(\vec{c}_{i,j}-\mat{H} \vec{x}_{i,j-1}\right)$ and $\mathrm{cov}(\vec{z}_{i,j}) = \mathrm{cov}(\mat{M} \vec{c}_{i,j} - \mat{F} \mat{M} \vec{c}_{i,j-1}) - \mathrm{cov}(\vec{w}_{i,j})$, where $\vec{K}_{i,j}$ is the constant Kalman gain and $\mat{M} = \left( \mat{H}^\top \mat{H} \right)^{-1} \mat{H}^\top$. We follow the approach described in \cite{basso2017kalman} to estimate $\mathrm{cov}(\vec{w}_{i,j})$ used in the above update. The update of $\mathrm{cov}(\vec{w}_{i,j})$ enables ImmTrack to adapt to dynamic sensing performance of the radar due to the position variations of users.

Note that the distance-based heuristic rule of transferring the PID of the initial centroids to the resulting centroids of the $k$-means clustering may have errors when the trajectories of two users cross in the radar's FoV. However, since the RKF is mainly used to assist better choosing the initial centroids rather than track the users, the swap of PIDs does not have long-lasting negative effect after the crossing because the models in Eq.~(\ref{eq:rkf}) are Markovian. Note that tracking the users is the subject of \sect\ref{subsubsec:cluster-tracking}.

\subsubsection{Evaluation}



\begin{figure}
  \begin{minipage}[t]{.6\linewidth}
    \begin{subfigure}[t]{\linewidth}
      \includegraphics[width=\linewidth]{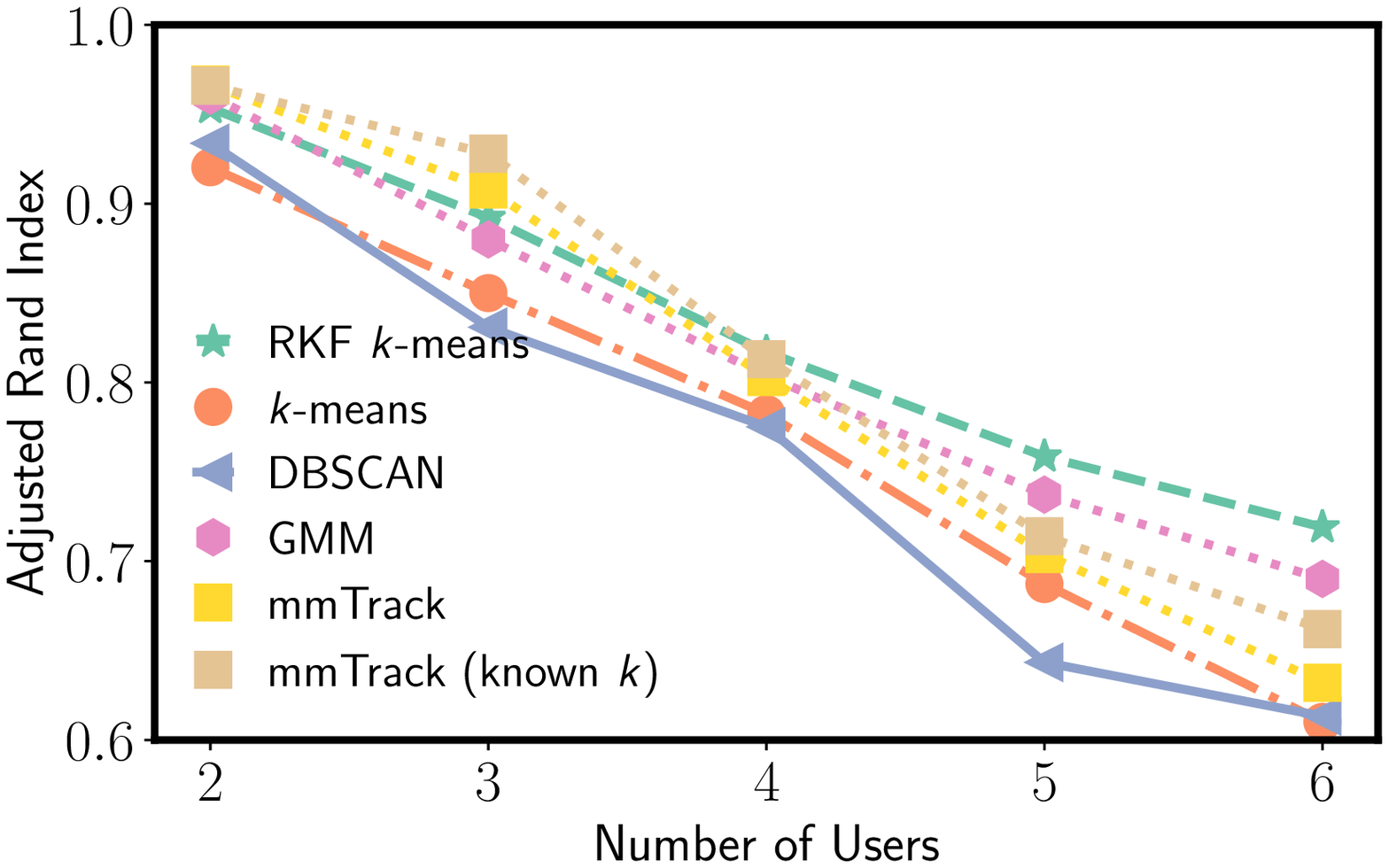}
      \caption{Performance of intra-frame clustering algorithms. Baselines: $k$-means, DBSCAN, GMM, mmTrack and its variant.}
      \label{fig:cluster algorithm}
    \end{subfigure}
  \end{minipage}
  \hfill
  \begin{minipage}[t]{0.35\linewidth}
    \begin{subfigure}[t]{\linewidth}
      \includegraphics[width=\linewidth]{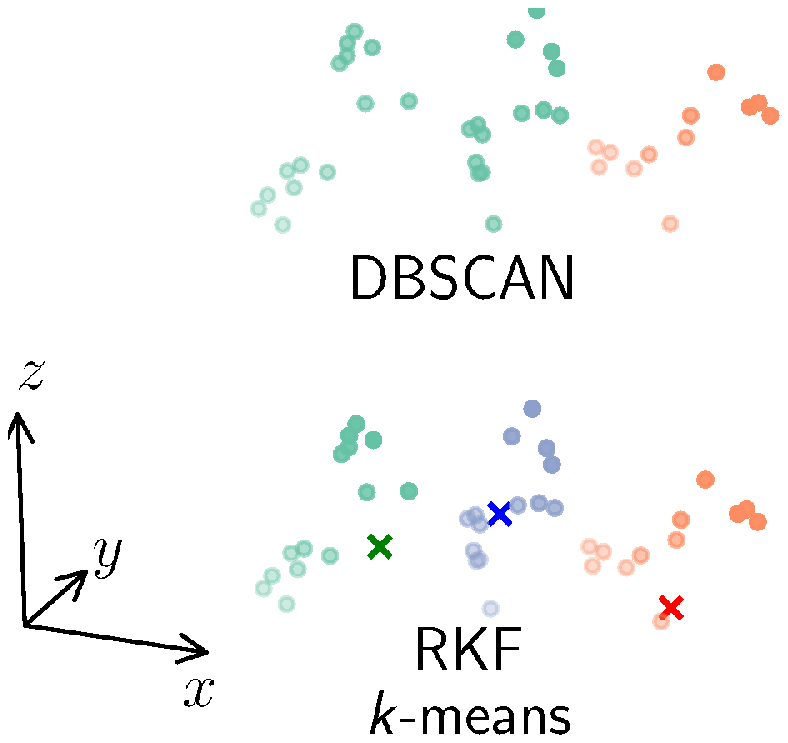}
      \caption{Results of RKF-assisted $k$-means and DBSCAN when $N=3$.}
      \label{fig:ClusterDemo}
    \end{subfigure}
  \end{minipage}
  \caption{Intra-frame clustering. The proposed RKF-assisted $k$-means clustering algorithm outperforms $k$-means, DBSCAN, and GMM. It also outperforms mmTrack \cite{wu2020mmtrack} when $N \ge 4$. In (b), color represents cluster ID, cross represents centroid, and DBSCAN yields 2 clusters for 3 users.}
  \vspace{-1em}
\end{figure}
  

We compare our RKF-assisted $k$-means algorithm with a variant without RKF and several other clustering approaches including DBSCAN and Gaussian mixture model (GMM) built with the expectation-maximization (EM) algorithm. We also implement the clustering algorithm proposed in mmTrack \cite{wu2020mmtrack}. The mmTrack applies the $k$-means algorithm with random initial centroids to cluster the point cloud. During the $k$-means iterations, mmTrack uses the medoids of the clusters obtained in the previous iteration as the initial centroids of the next iteration. The mmTrack determines the value of $k$ using the silhouette analysis. In addition, we implement a variant of mmTrack's clustering algorithm by removing the silhouette analysis and directly setting $k=N$. All the above baseline approaches do not consider motion.



We use the Adjusted Rand Index (ARI) to measure the quality of clustering. Zero ARI indicates random guessing-like clustering, whereas ARI of one suggests perfect clustering.
We compute per-frame ARIs and report the average ARI.
During the experiment, the users follow pre-defined trajectories, so that we can obtain the ground truth. More details of the experiment setup are presented in \sect\ref{sec:eval}. From Fig.~\ref{fig:cluster algorithm},
our RKF-assisted $k$-means outperforms $k$-means, DBSCAN, and GMM. When $N \le 3$, the mmTrack and its variant with known $k$ slightly outperform our RKF-assisted $k$-means in terms of ARI. However, the advantage of our RFK-assisted $k$-means over mmTrack and its variant increases with $N$ when $N \ge 4$. \yimin{The explanations for the above results are as follows. When the occlusion cases increase due to the increase of users, our RKF-assisted $k$-means algorithm outperforms mmTrack. When there are no or limited occlusions, mmTrack's clustering algorithm performs well. However, with our RKF-assisted $k$-means algorithm, some of the points corresponding to users in the point cloud are excluded in the phase of static points removal, leading to lower ARI.}
Fig.~\ref{fig:ClusterDemo} shows the clustering results of the DBSCAN and RKF-assisted $k$-means algorithms when $N=3$, respectively. DBSCAN mistakenly combines two users into a single cluster. The above results suggest that the consideration of motion improves clustering performance.

\subsection{IMU-Assisted Inter-Frame Cluster Tracking}
\label{subsubsec:cluster-tracking}


\subsubsection{Design}
The association of the clusters in the consecutive frames that correspond to the same user is based on {\em space coherence} and {\em motion coherence}. The former means that the shape of a moving object at close locations are similar from the radar's perspective; the latter means that the object's motions in consecutive frames are similar. We design a new deep learning-based feature extractor called {\em mmClusterNet} that fuses {\em shape}, {\em motion}, and {\em IMU PID} features of a cluster into a single {\em cluster tracking feature} for each frame.






\begin{figure}
  \centering
  \includegraphics[width=\linewidth]{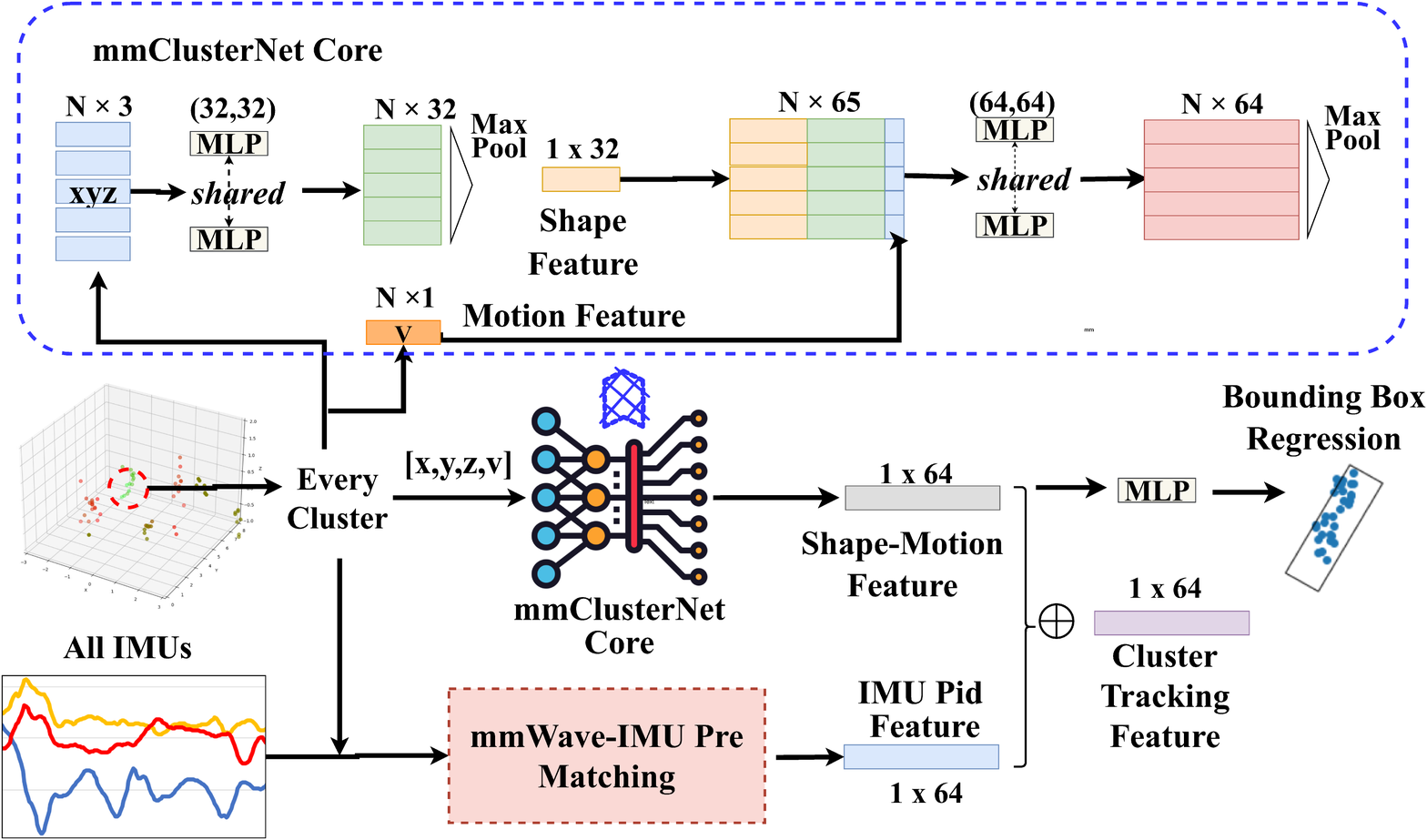}
  \vspace{-2em}
  \caption{mmClusterNet for fusing shape, motion, IMU PID features. The distance matrix of fused feature of clusters is used as the metric for inter-frame association and tracking. }
  \label{fig:feature_extractor}
  \vspace{-1em}
\end{figure}



Fig.~\ref{fig:feature_extractor} shows mmClusterNet's architecture. For each frame, it takes each of the clusters produced by \sect\ref{subsec:mmWave-track-cluster} as input. The mmClusterNet is designed to process a cluster with $n$ 3D points, where $n$ is fixed at the design phase. When processing a smaller cluster, ImmTrack firstly applies interpolation to generate an $n$-point cluster. For the AWR1843 mmWave radar, $n=24$ is a good setting because it is an empirical upper bound of human cluster size. 
As shown in the upper branch in Fig.~\ref{fig:feature_extractor}, each of the $n$ points is processed by a shared multilayer perceptron (MLP) with two 32-neuron hidden layers. The results of the $n$ shared MLPs are max-pooled to generate a $1\times 32$ shape feature, which is copied vertically $n$ times, concatenated with the shared MLPs' outputs and the cluster's radial velocity vector (as the motion feature) to form an $n \times 65$ tensor. Then, each row of the tensor is processed by an MLP with two 64-neuron hidden layers and max-pooling to produce a $1 \times 64$ shape-motion feature. Finally, the shape-motion feature is fused with the IMU PID feature, which is detailed shortly, by element-wise addition to produce the cluster tracking feature. To train mmClusterNet, we append a regression MLP as the downstream task that produces a bounding box of the cluster from the shape-motion feature. Then, we use manually labeled bounding boxes as ground truth to train the mmClusterNet core. In \sect\ref{compare_down_task}, we will evaluate the impact of different choices of the downstream task on mmClusterNet's performance. \yimin{Note that the training data for the mmClusterNet core is unnecessary to be {\em in situ} data. The training can be based on a public dataset such as ShapeNet.}

The shape-motion feature is directly affected by the radar's sensing noises. Thus, we supplement user-specific static information (i.e., the IMU PID feature) to assist the cluster tracking. Specifically, we perform a {\em pre-matching} between the clusters generated by the radar and the IMUs, and then use the matched IMU's PID as the user-specific static information. The pre-matching is as follows. First, we compute $\overline{\vec{v}}_i$, which is the weighted average 3D velocity of all points in the $i^\text{th}$ cluster, by $\overline{\vec{v}}_i= \sum_{s=1}^{n_i} \frac{\textit{projection}_{\vec{v}_{\vec{c}_i}}\vec{v}_{i,s}}{\ln n_i \cdot \textit{rank}(d_s, \{d_1, \ldots, d_{n_i}\})}$, where
$n_{i}$ is the number of points in the cluster, $\vec{c}_i$ is the cluster centroid, $\vec{v}_{\vec{c}_i}$ is $\vec{c}_i$'s 3D velocity, $\vec{v}_{i,s}$ is the 3D velocity of the $s^\text{th}$ point of the cluster, $d_s$ is the Euclidean distance between the $s^\text{th}$ point and the centroid, the operator $\textit{projection}_\vec{a} \vec{b}$ returns the projection of $\vec{b}$ in the direction of $\vec{a}$, and the operator $\textit{rank}(a, A) \in \{1, \ldots, |A|\}$ returns the rank of $a$ in the set $A$ with elements in ascending order. With the reciprocal of rank as the weight, a point closer to the centroid receives a larger weight in the averaging. Using the rank instead of distance as weight for velocity avoids the issue of physical unit conciliation.
We apply the coefficient $\frac{1}{\ln n_i}$ to make the sum of the weights to be approximately one, i.e., $\sum_{s=1}^{n_i} \frac{1}{\ln n_i \cdot \textit{rank}(d_s, \{d_1, \ldots, d_{n_i}\})} \approx 1$. Second, with all clusters' average velocity magnitudes $\{|\overline{\vec{v}}_1|, \ldots, |\overline{\vec{v}}_{N}|\}$ and all IMUs' velocity magnitudes denoted by $\{|\vec{u}_1|, \ldots, |\vec{u}_{N}|\}$, we apply the Hungarian algorithm to find the one-to-one pre-match between the clusters and IMUs based on Euclidean distance. Let $\mathrm{PID}_i \in \{1, \ldots, N\}$ denote the PID of the IMU pre-matched with the $i^\text{th}$ cluster. We apply the position encoding \cite{vaswani2017attention} to generate the $i^\text{th}$ cluster's $1 \times 64$ IMU PID feature as $[g_1, h_1, g_2, h_2, \ldots, g_{32}, h_{32}]$, where $g_m =\sin \left( \left( \frac{\mathrm{PID}_i}{1000}\right)^{\frac{m}{32}} \right)$ and $h_m=\cos \left( \left( \frac{\mathrm{PID}_i}{1000}\right)^{\frac{m}{32}} \right)$. As presented earlier, the IMU PID feature is added to the shape-motion feature to form the cluster tracking feature.

Given the cluster tracking features obtained in two consecutive frames, the Hungarian algorithm is used to associate one feature in the former frame and one feature in the latter, exclusively, based on cosine similarity. The associated clusters are considered from the same user. In addition, their centroids over time form the trajectory of the user. All trajectories will be input to the trajectory-based association module presented in \sect\ref{subsec:association}.

\begin{figure}
  \centering
  \begin{subfigure}[b]{.23\linewidth}
    \centering
    \includegraphics[width=\linewidth]{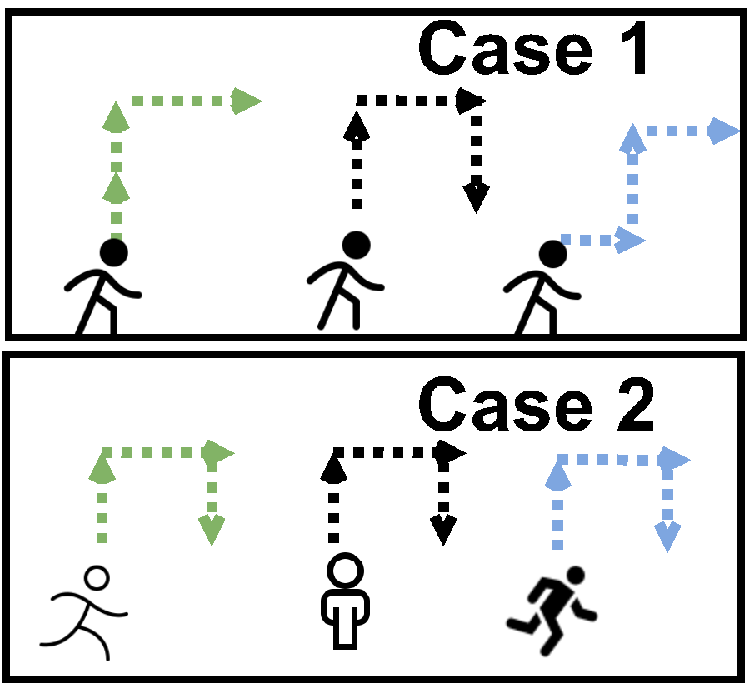}
    \caption{Setup.}
  \end{subfigure}%
  \hfill
  \begin{subfigure}[b]{.37\linewidth}
    \centering
    \includegraphics[width=\linewidth]{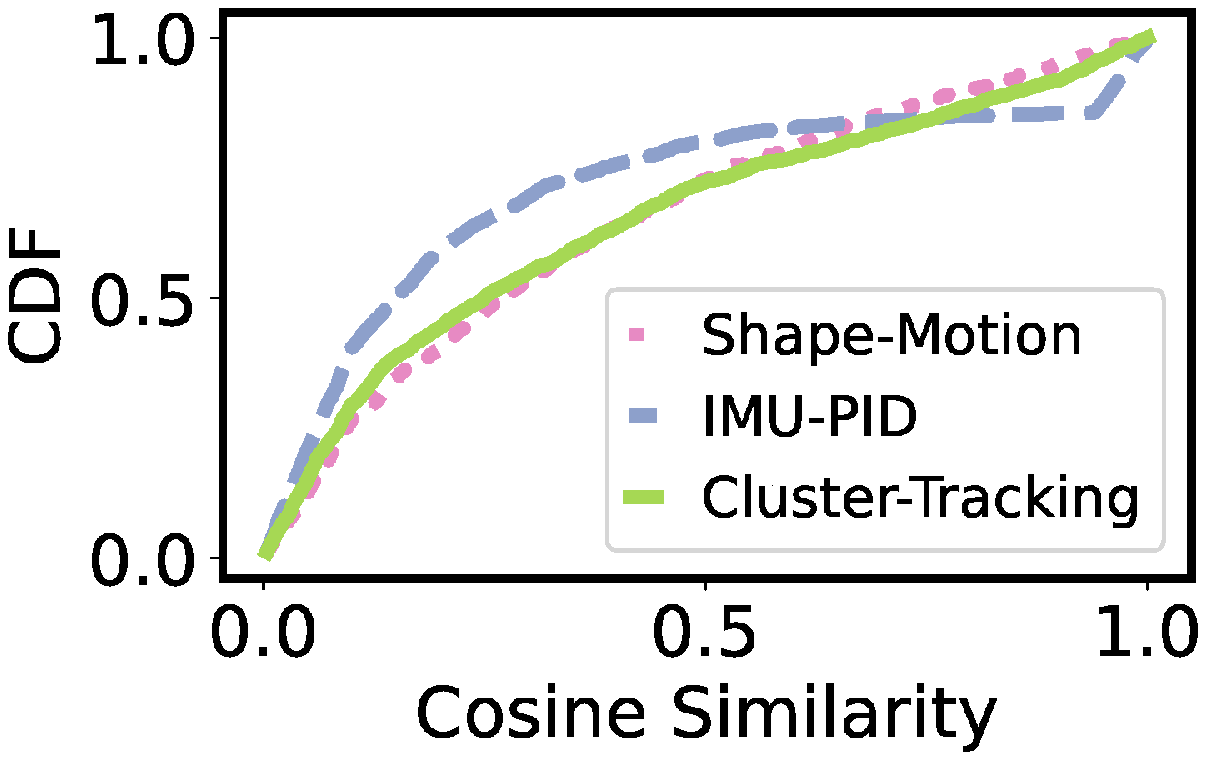}
    \caption{Case 1 result.}
  \end{subfigure}
  \hfill
  \begin{subfigure}[b]{.37\linewidth}
    \centering
    \includegraphics[width=\linewidth]{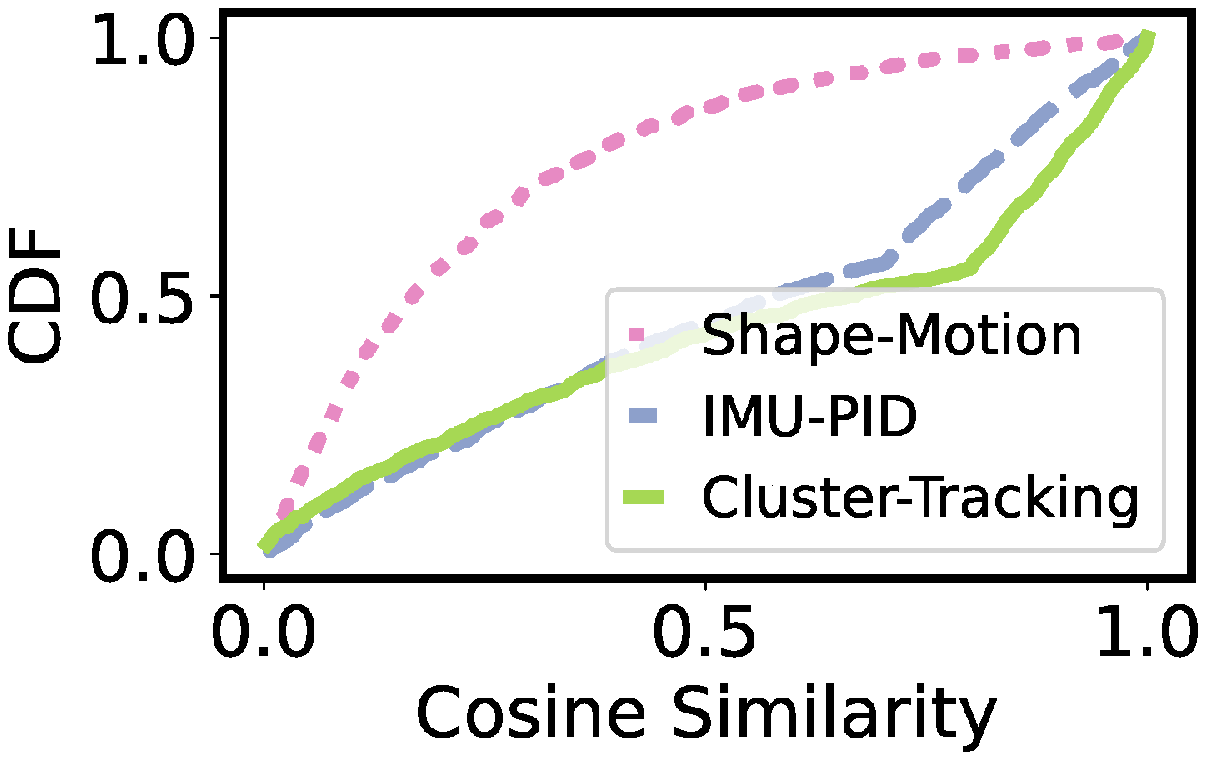}
    \caption{Case 2 result.}
  \end{subfigure}
  \vspace{-1em}
  \caption{Cosine similarity between features of clusters of same user in two consecutive frames.}
  \label{fig:imuencoding}
  \vspace{-1em}
  \end{figure}

\subsubsection{Evaluation}
We evaluate the advantage of the cluster tracking feature, compared with solely using either shape-motion feature or IMU PID feature.
We consider two cases as illustrated in Fig.~\ref{fig:imuencoding}: (1) all users walk at the same speed but follow different paths of different shapes; (2) all users walk at different speeds and follow different paths of the same shape. We measure the cosine similarity between the features of the clusters corresponding to the same user in two consecutive frames.
Fig.~\ref{fig:imuencoding} shows the cumulative distribution functions (CDFs) of the measured cosine similarities in the two cases. 
In case (1), the performance of shape-motion feature is similar to cluster tracking feature. In case (2), the performance of IMU PID feature is similar to cluster tracking feature, because the velocity-based mmWave-IMU pre-matching is accurate when users' speeds are different and the matched IMU PID contributes more information than the shape-motion feature. The above results show that the cluster tracking feature takes both the advantages of shape-motion feature and IMU PID feature.

\section{Learning-Based Cross-Modality Trajectory Association}
\label{subsec:association}

\subsection{Design Principle}

This module identifies the correspondence among the trajectories reconstructed by the radar in \sect\ref{sec:global-tracking-design} and IMUs via dead reckoning, to re-identify radar's sensing results. Essentially, it is a weighted bipartite matching problem with trajectory similarity as the weight.
For association, we use the 2D trajectory (without including the altitude dimension), as it is a common feature that can be derived from both the radar's and IMUs' results and is agnostic to modality-dependent details. 
For either a radar cluster or an IMU, a trajectory over an {\em association time window} $[t_0, t_1]$ is denoted by $\mathcal{T}(t) = \{x(t), y(t) | t \in [t_0, t_1]\}$. To compute the similarity between a radar cluster's trajectory $\mathcal{T}_{r}(t)$ and an IMU's trajectory $\mathcal{T}_{i}(t)$, the radar's and IMU's 2D coordinate systems need to be registered. A potential method to register the two coordinate systems, both originating at the start points of $\mathcal{T}_r(t)$ and $\mathcal{T}_i(t)$, is to exhaustively search a relative angle between them such that the similarity between $\mathcal{T}_r(t)$ and $\mathcal{T}_i(t)$ under the candidate registration is maximized. However, this registration incurs high compute overhead.

We design a learning-based, registration-free association approach. 
The main idea is that, instead of considering the distance between two registered trajectories in the same Euclidean space, we take advantage of the feature extraction capability of neural networks to transform trajectories into high-dimensional features, and perform the association based on the distance in the high-dimensional space.
Specifically, we first encode the trajectory into an imagery representation, called {\em trace map}. This is a preparation step that restructures data to a uniform and compact form.
Then, we feed {\em trace maps} from the two modalities into a Siamese neural network for feature extraction, based on whose outputs, the distance matrix can be calculated. 
Finally, in association, we introduce a soft voting mechanism which aggregates the information of multiple association time windows and thus mitigates the short-time interference. 
To train the Siamese network, we do not require the ground-truth trajectories. Instead, we extensively construct positive pairs and negative pairs of trajectories, and use a triplet loss to push negative pairs away while bringing together positive pairs, where the only labels required are the matching relationships of the trajectories from the two modalities.

\subsection{Trace Map Generation}

Let $\mathcal{M}=\{M(x,y) | \forall (x,y)\}$ denote a trace map converted from a trajectory $\mathcal{T}(t)$, where the pixel value $\mathcal{M}(x,y)$ encodes all the times elapsed from when the trajectory crosses the location $(x,y)$. Let $f_s$ denote the sampling rate in frames per second (fps) of the sensor. Let $T(x,y)$ denote the set of the time instants at which the trajectory crosses $(x,y)$. If $T(x,y) \neq \emptyset$, the map pixel value is given by $\mathcal{M}(x,y) = \sum_{t \in T(x,y)} f_s \cdot (t - t_0)$, where $t_0$ denotes the time instant that the trajectory starts; otherwise, $\mathcal{M}(x,y) = 0$. 
Intuitively, $\mathcal{M}(x,y)$ encodes the number of frames passed when the user's trajectory crossed $(x, y)$ since the trajectory begins. Then, ImmTrack converts the obtained trace map into an image with three 8-bit channels of RGB data.
We use $\mathcal{M}_r$ and $\mathcal{M}_i$ to denote the color trace maps converted from $\mathcal{T}_r(t)$ and $\mathcal{T}_i(t)$, respectively.


\begin{figure}
\centering
\begin{subfigure}{.15\textwidth}
    \centering
    \includegraphics[width=\linewidth]{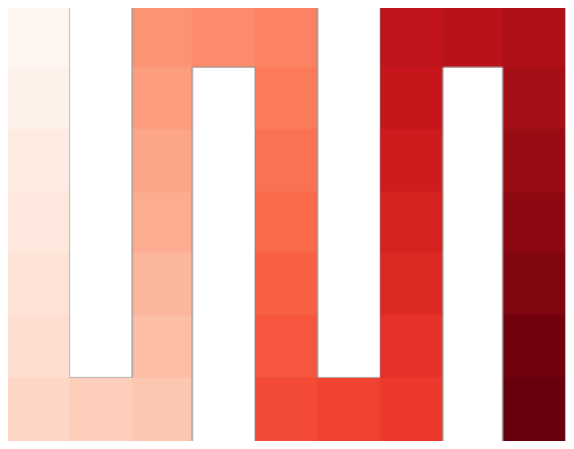}  
    \caption{truth ($\rho$=0.5)}
    \label{figure:gt5}
\end{subfigure}
\begin{subfigure}{.15\textwidth}
    \centering
    \includegraphics[width=\linewidth]{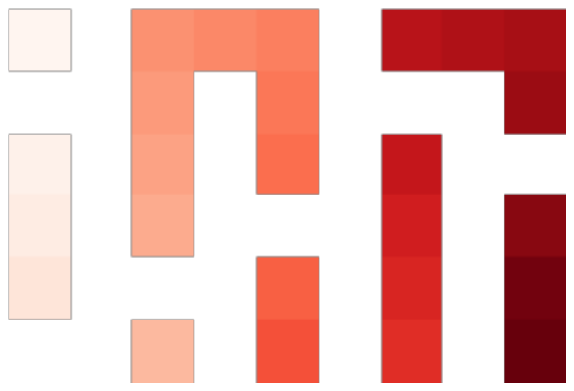}  
    \caption{radar ($\rho$=0.5)}
    \label{figure:mmwave5}
\end{subfigure}
\begin{subfigure}{.15\textwidth}
    \centering
    \includegraphics[width=\linewidth]{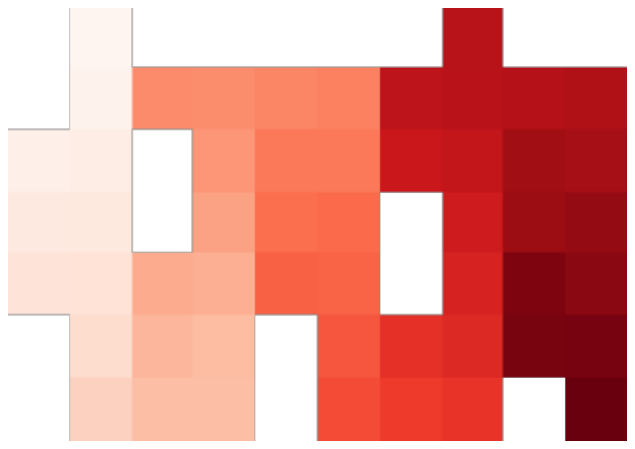}  
    \caption{IMU ($\rho$=0.5)}
    \label{figure:imu5}
  \end{subfigure}
  \hfill
\begin{subfigure}{.15\textwidth}
    \centering
    \includegraphics[width=\linewidth]{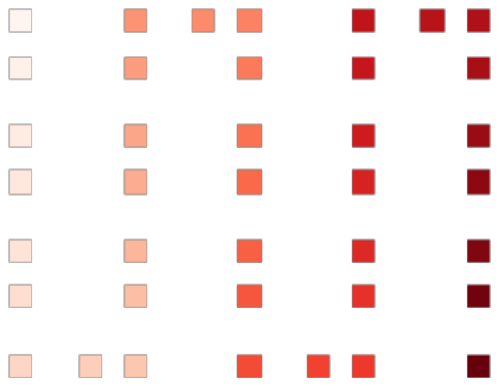}  
    \caption{truth ($\rho$=0.2)}
    \label{figure:gt2}
\end{subfigure}
\begin{subfigure}{.15\textwidth}
  \centering
  \includegraphics[width=\linewidth]{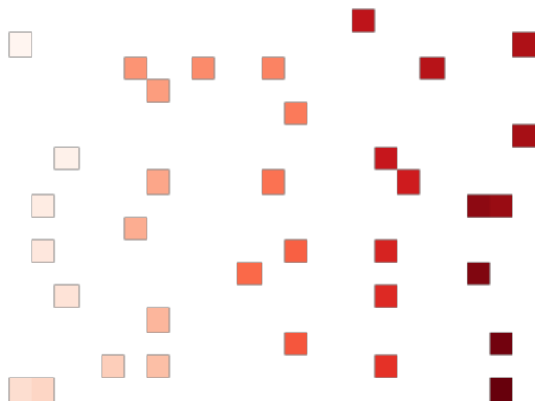}   
  \caption{radar ($\rho$=0.2)}
  \label{figure:mmwave2}
\end{subfigure}
\begin{subfigure}{.15\textwidth}
  \centering
  \includegraphics[width=\linewidth]{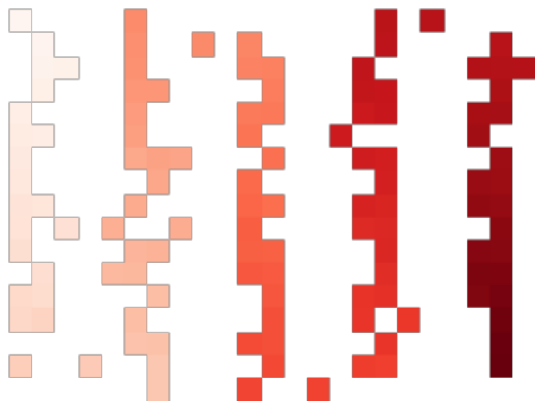}  
  \caption{IMU ($\rho$=0.2)}
  \label{figure:imu2}
\end{subfigure}
\vspace{-1em}
  \caption{Trace maps of ground truth, radar, IMU trajectories.}
  \label{fig:demoOnS}
\end{figure}

Furthermore, in order to mitigate the impact of noises, we adopt specific spatial grid size $\rho$ for the trace maps.
Fig.~\ref{fig:demoOnS} shows the trace maps of the ground truth, radar, and IMU trajectories under two $\rho$ settings, where a user follows a square zig-zag path to move. A darker red pixel indicates that the trajectory crosses the position more recently. We can see that, due to the inherent uncertainty of sensing, the radar's and IMU's trace maps have deviations from the ground truth. Moreover, under a certain $\rho$ setting, the IMU's trace map has more colored pixels on the trace than the radar's because of IMU's higher sampling rate. As a result, for IMU, setting a smaller $\rho$ can better reduce the crosstalks among different segments of the trajectory, while a larger $\rho$ can make the trace for the radar more continuous. 
In the rest of this paper, we adopt $\rho=0.2\,\text{m}$ and $\rho=0.5\,\text{m}$ for IMU and radar, respectively.
Finally, we crop the trace map in an area of 20m $\times$ 20m and resize it to 193 $\times$ 193, which will be fed into the Siamese neural network presented in \sect\ref{subsec:siamese}.
\subsection{Comparative Features Extraction}
\label{subsec:siamese}

\begin{figure}
  \centering
  \begin{subfigure}[]{0.28\linewidth}
    \centering
    \includegraphics[width=\linewidth]{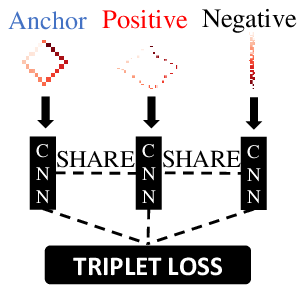}
    \label{fig:siamese_general}
  \end{subfigure}
  \hspace{0.45em}
  \begin{subfigure}[]{0.68\linewidth}
    \includegraphics[width=\linewidth]{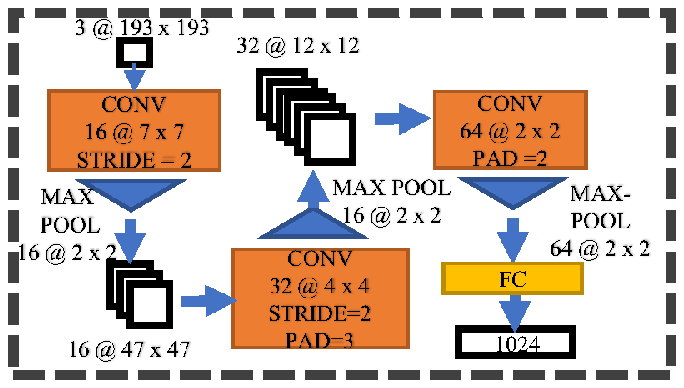}    
    \label{fig:siamese_cnn}    
  \end{subfigure}
  \vspace{-3em}
  \caption{Left: Siamese network using three identical CNNs with shared weights during training. Right: Architecture of CNN that extracts comparative feature from trace map.}
  \label{fig:siamese}
\end{figure}

We design a Siamese neural network to extract comparative features from $\mathcal{M}_r$ and $\mathcal{M}_i$, whose cosine similarity characterizes how close the $\mathcal{T}_r(t)$ and $\mathcal{T}_i(t)$ are.
Typically, a Siamese network contains two or more identical sub-networks that extract features from their respective input. During training, any parameter updates are mirrored across all sub-networks.
As illustrated in Fig.~\ref{fig:siamese}, the Siamese network used by ImmTrack employs a convolutional neural network (CNN) as the feature extractor. The CNN consists of three convolutional layers with rectified linear unit (ReLU) activation followed by max-pooling and a final fully-connected layer producing a $1 \times 1024$ feature vector.
During training, three such identical CNNs are used to process three inputs, i.e., anchor, positive, and negative inputs.
The anchor and positive inputs are two trace maps generated from the radar and IMU for the same user at the same time, while the negative input is an unrelated trace map from either the radar or IMU.
Denoting by $\vec{f}_a$, $\vec{f}_p$, and $\vec{f}_n$ the feature vectors produced by the CNN for the anchor, positive, and negative inputs, we use the triplet loss function for training: $\mathcal{L}=\max ( \| \vec{f}_a - \vec{f}_p\|_{\ell_2}- \| \vec{f}_a - \vec{f}_n \|_{\ell_2} + \mathrm{margin}, 0)$.
We also generate simulated trajectories to augment \yimin{the training data collected in a real environment}. 
Specifically, we use a random walk stochastic process to generate the anchor, and obtain the positive input by scaling up or down the anchor and shifting $10\%$ of the anchor positions to their neighbors. \yimin{Note that the training data needed by the Siamese neural network is unnecessary to be {\em in situ} data, because the network only learns extracting environment-agnostic comparative features.}
At ImmTrack's run time, the trained CNN is used to extract the comparative feature from any given trace map $\mathcal{M}$.



\subsection{Cross-Modality Association}\label{oneshot_id_ass}  

For the $w^\text{th}$ time step in an association time window, 
ImmTrack constructs a similarity matrix $\mat{S}_w \in \mathbb{R}^{N \times N}$, where its $(i,j)^\text{th}$ element is the cosine similarity between the comparative feature vectors extracted by the Siamese network from the trace maps of the $i^\text{th}$ radar cluster and $j^\text{th}$ IMU, respectively.
ImmTrack generates an average similarity matrix, denoted by $\mat{S}$, over a total of $W$ consecutive association time windows, i.e., $\mat{S}=\frac{1}{W} \sum_{w=1}^{W}\mat{S}_w$. Hungarian algorithm is applied to propose an association between the radar clusters and IMUs. If the proposal is accepted, the IMUs' PIDs are transferred to the radar clusters for re-identification.
\sect\ref{main_acc_result} will show via evaluation that the multi-window similarity averaging improves the robustness of the association, compared with using a single window only.

In addition, ImmTrack applies two criteria to accept an association proposal. If either criterion is not met, ImmTrack excludes the oldest window from the $W$ windows, waits for a new window becoming available, and checks the two criteria again.
The two criteria are as follows. {\bf Criterion~1:} For each pair of associated radar cluster and IMU, the similarity between their comparative features needs to be higher than a pre-defined threshold $\alpha$. This criterion sets a lower bound for the association quality.
The $\alpha$ can be set according to the data used to train the Siamese network by $\alpha = \max \{\min_{\forall (\vec{a}, \vec{p}) \in \mathcal{P} } S_c(\vec{a},\vec{p}), \max_{\forall (\vec{a}, \vec{n}) \in \mathcal{N}} S_c(\vec{a},\vec{n}) \}$, where $\mathcal{P}$ and $\mathcal{N}$ are the positive and negative pair sets, $S_c(\cdot, \cdot)$ denotes cosine similarity. Our training data gives $\alpha=0.23$.
{\bf Criterion~2:} Any IMU cannot produce the highest cosine similarity with two or more radar clusters among all IMUs. Formally, $\forall i \in [1, N]$, if the $(i,j)^\text{th}$ element of $\mat{S}$ (denoted by $\mat{S}_{i,j}$) is the maximum value within the $i^\text{th}$ row of $\mat{S}$, then $\nexists k \in [1,N]$ such that $\mat{S}_{k,j}$ is the maximum value within the $k^\text{th}$ row of $\mat{S}$. This criterion makes sure that the IMU most similar with every radar cluster is unique.

\subsection{Handling Users with Identical Trace Maps}
\label{boundary_case_macth}

Multiple users may generate nearly identical trace maps in certain cases, e.g., when they walk side by side or follow simple straight paths.
Within a certain modality, such nearly identical trace maps can be detected by checking their pair-wise similarities.
\yimin{Based on a dataset collected from six human subjects in controlled experiments with pairs of human subjects walking side by side, the detection rates of identifying the side-by-side walk are 92.5\% and 77.5\% using mmWave radar data and IMU data, respectively, by adopting a threshold of 0.92 on the normalized similarity for the detection. After removing the entries of the $\mat{S}_w$ corresponding to the detected identical trace maps, the remaining entries are processed by the cross-modality association presented in \sect\ref{oneshot_id_ass}.} This section presents a separate cross-modality association approach for the nearly identical trace maps based on gait analysis. \yimin{ImmTrack initializes the gait analysis if it detects users with nearly identical trace maps from the mmWave radar.} The gait analysis for an mmWave cluster is as follows. First, we compute the measured spectrogram $\mat{X}_m(v_k, t_l)$ from the Doppler Fourier transform corresponding to the points belonging to the cluster, where $v_k$ and $t_l$ represent the velocity and time bins, respectively. Second, we use the Boulic model \cite{boulic1990global} to generate the simulated spectrogram $\mat{X}_s(v_k, t_k | f_c, l_c, \varphi_c)$, where the parameters $f_c$, $l_c$, and $\varphi_c$ are the specified step frequency, step length, and start phase, respectively. \yimin{By solving $
\mathop{\arg\min}_{f_c, l_c, \varphi_c}
\sum_{\forall v_k, t_l} \left\| \mat{X}_{\mathrm{m}}^{\log }\left(v_k, t_l\right) 
- \mat{X}_{\mathrm{s}}^{\log }\left(v_k, t_l | f_c, l_c, \varphi_c \right) \right\|_{\ell_2}^2
$, where the superscript ``log'' means element-wise log normalization, the gait feature $(f_c, l_c)$  is estimated from mmWave radar data.}
For IMU data, we employ the IMU-based gait analysis \cite{madgwick2011Imutrack} to estimate the gait feature $(f_c, l_c)$. Lastly, Hungarian algorithm is applied to associate the mmWave clusters and IMU traces that respectively produce nearly identical trace maps, in terms of the cosine similarity between the mmWave-based and IMU-based gait features. The effectiveness of the mechanism presented in this section will be evaluated in \sect\ref{eval_tiny_space}.

\section{Implementation and Evaluation}
\label{sec:eval}

\begin{figure}
\centering
\begin{subfigure}[t]{.3\linewidth}
  \centering
  \includegraphics[width=\linewidth]{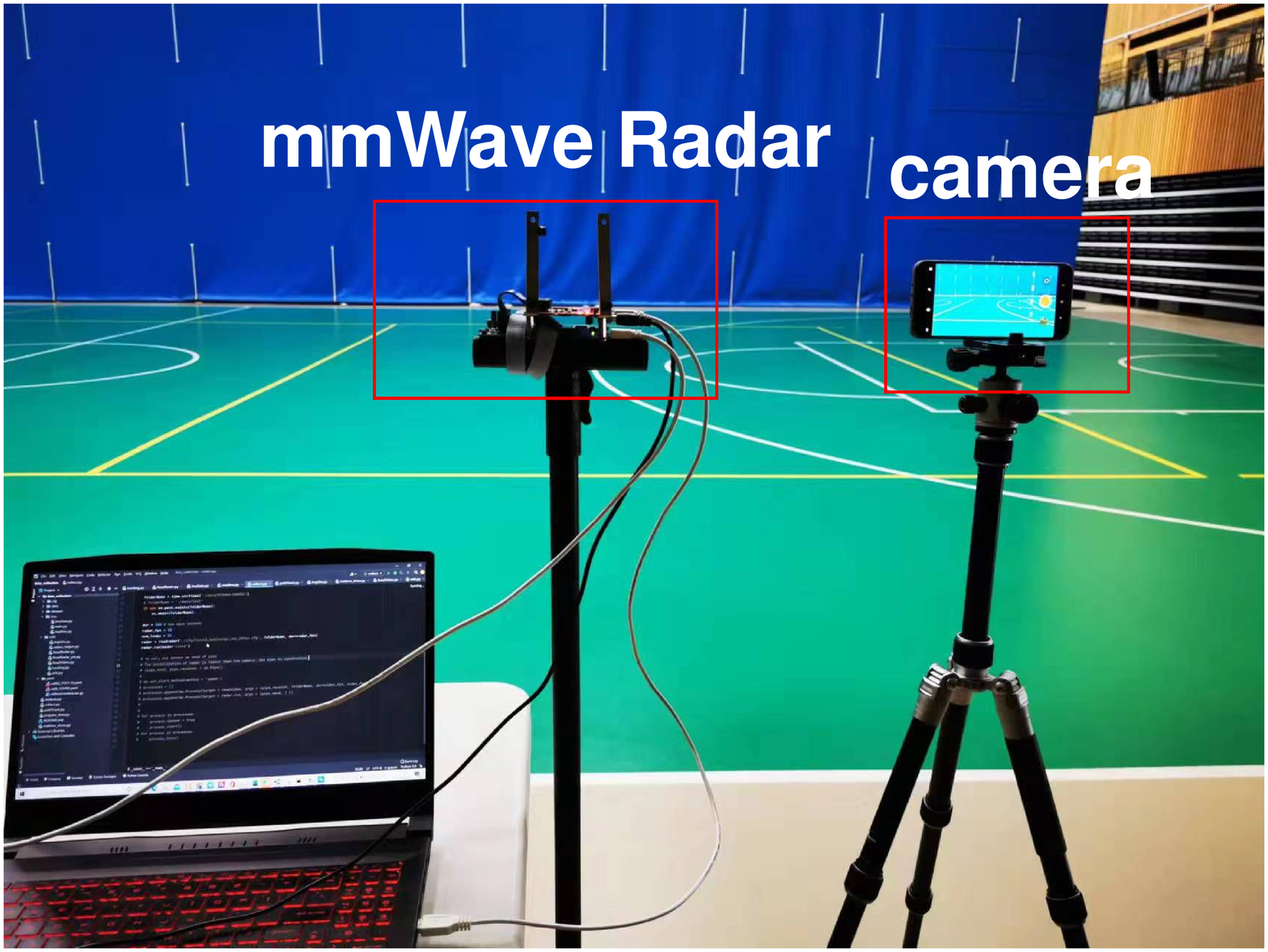}
  \caption{Sports hall setup}
\end{subfigure}%
\hfill
\begin{subfigure}[t]{.3\linewidth}
  \centering
  \includegraphics[width=\linewidth]{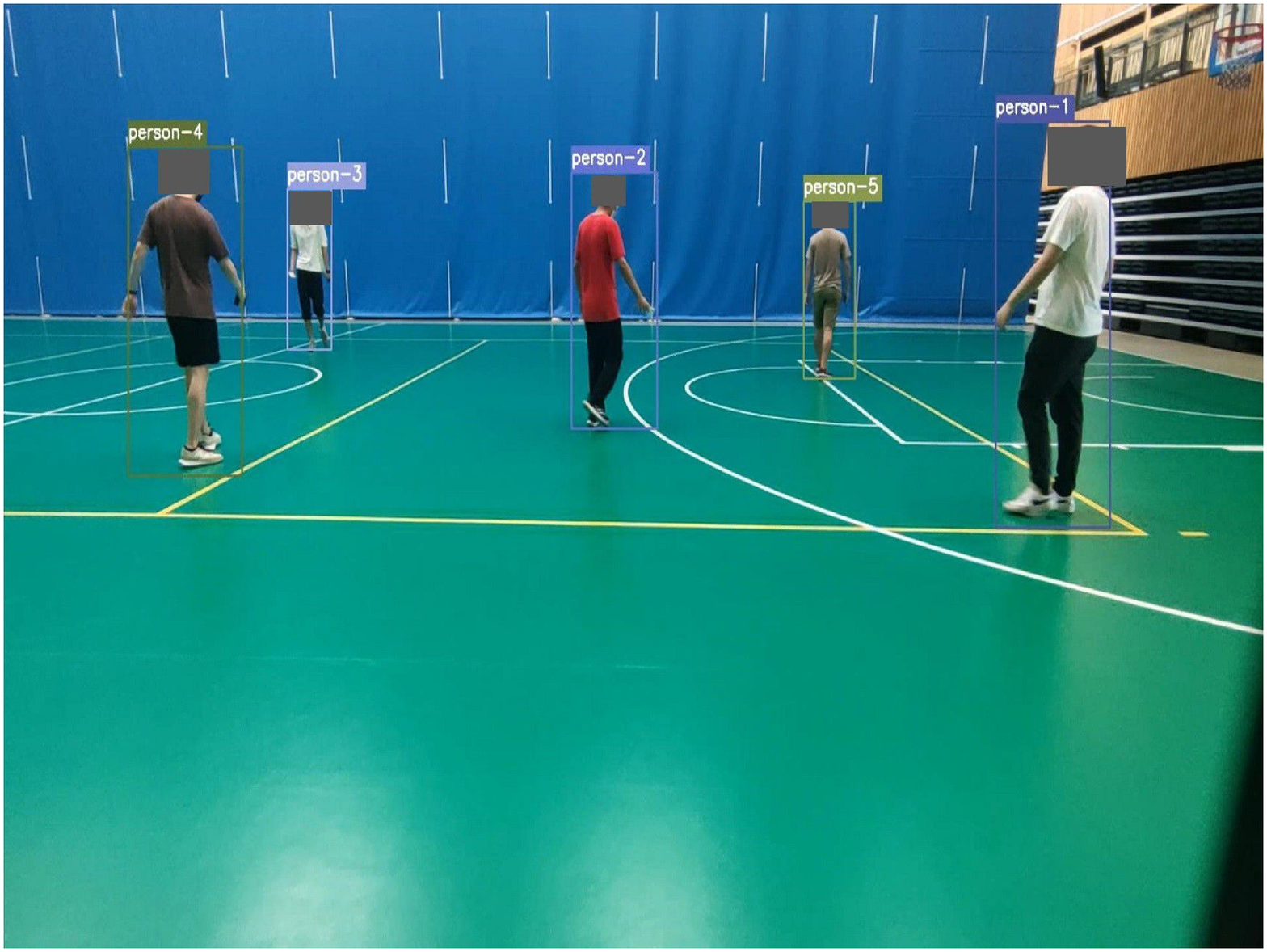}
  \caption{Subject detection}
  \label{fig:subject-detection}
\end{subfigure}
\hfill
\begin{subfigure}[t]{.3\linewidth}
  \centering
  \includegraphics[width=\linewidth]{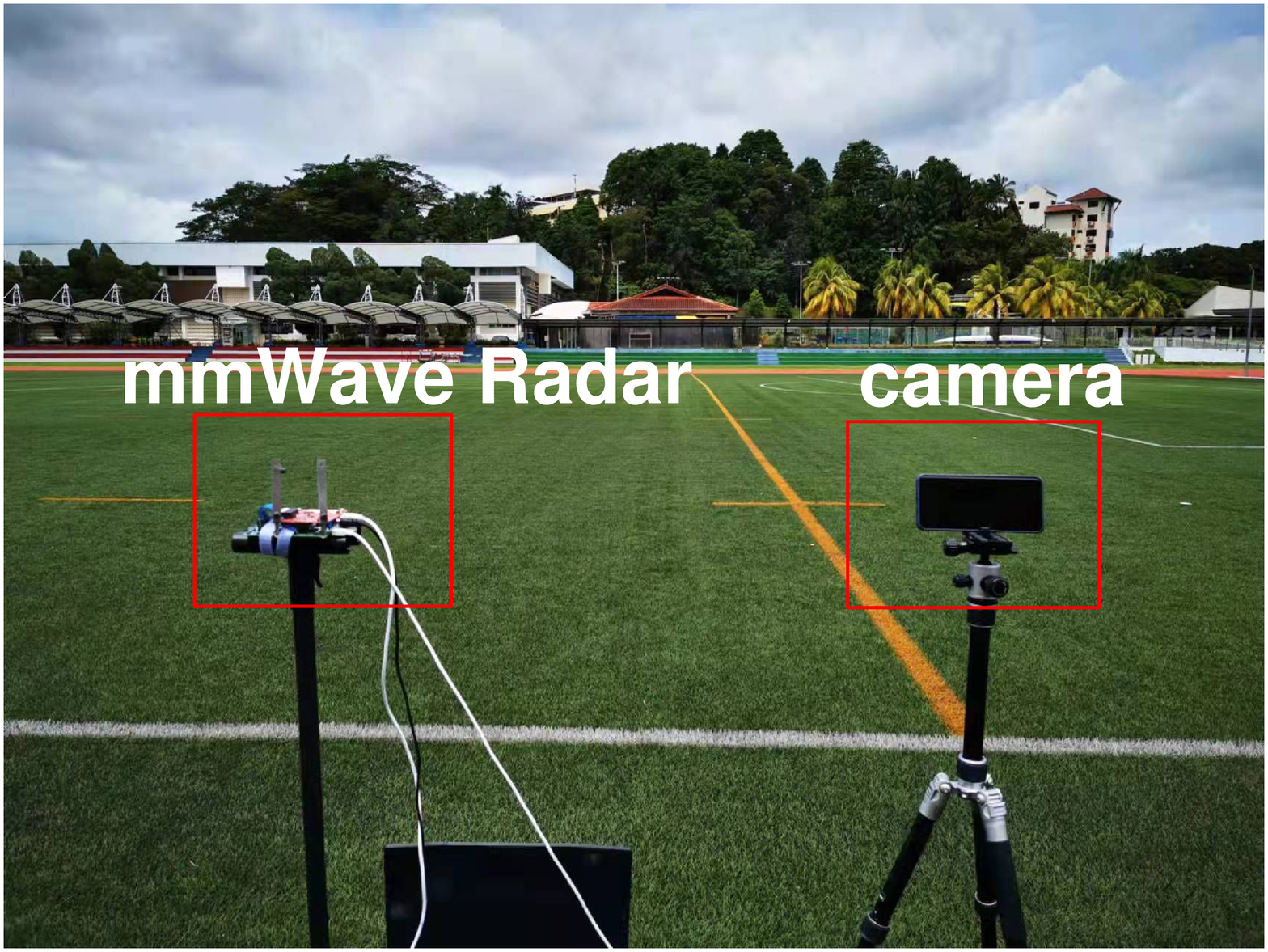}
  \caption{Outdoor setup}
\end{subfigure}
\vspace{-1em}
\caption{The sports hall and outdoor experiment setups.}
\label{fig:different scene}
\vspace{-1em}
\end{figure}

We have implemented ImmTrack using a Texas Instrument AWR1843 mmWave radar hosted by a laptop computer. The users use their own smartphones of various models to participate in the evaluation.\footnote{Volunteers' participation is under NTU IRB protocol with reference no. IRB-2022-309.} We collect IMU data using the MATLAB Mobile app running on the users' smartphones. The sampling rates of the radar and IMU are $8\,\text{fps}$ and $100\,\text{fps}$, respectively. 
The association time window is 12 seconds, with 2-second overlap between two consecutive windows.
For cross-modality association, we set $W=3$, i.e., the similarity matrices in three consecutive association windows are averaged.
We primarily conduct experiments in an indoor sports hall and an outdoor space as shown in Fig.~\ref{fig:different scene}.
We also conduct experiments in a lab space as shown in Fig.~\ref{Fig:floorPlan} with up to 27 people.

\subsection{Cross-Modality Association Performance}\label{system_performance}



\begin{figure}
  \begin{subfigure}[t]{\linewidth}
    \centering
    \includegraphics[width=\linewidth]{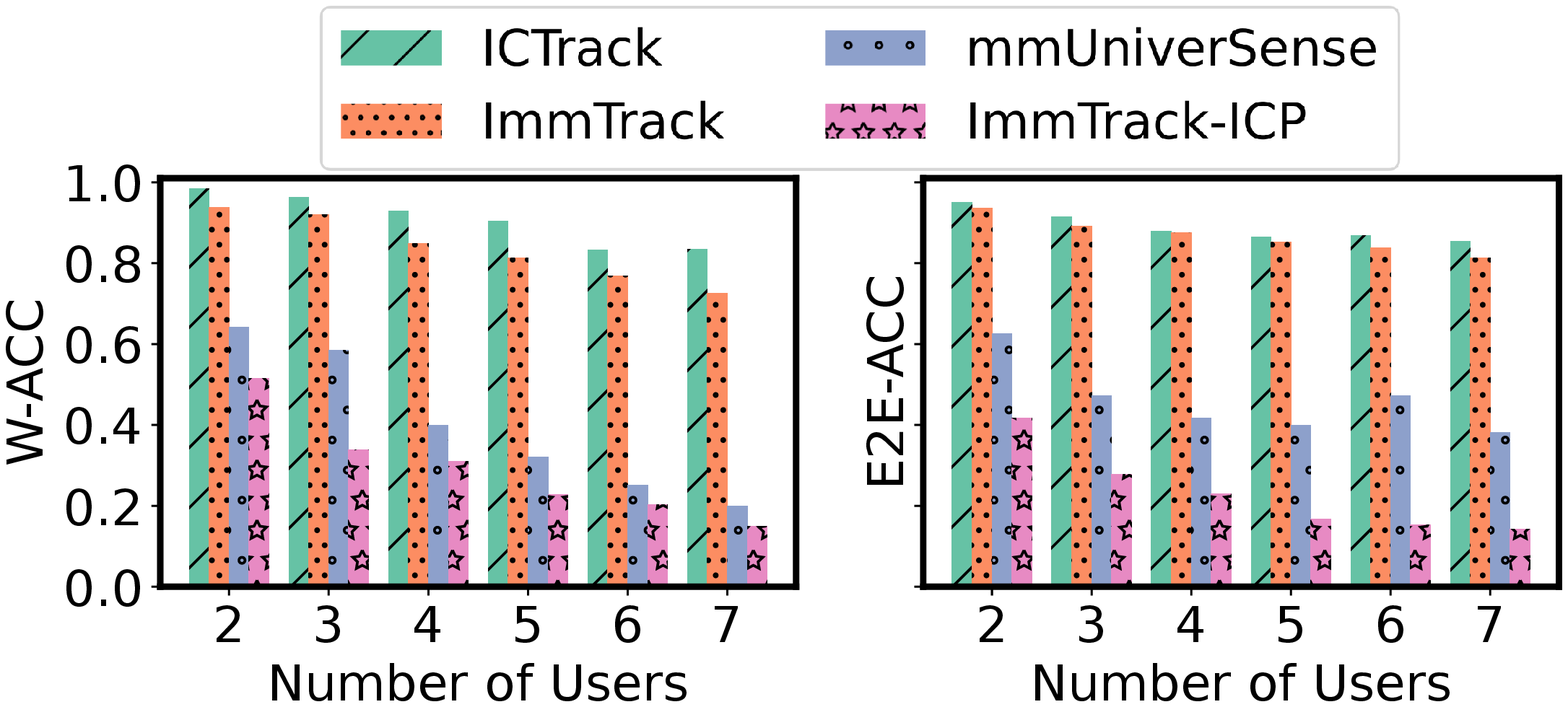}
   \caption{Overall W-ACC and E2E-ACC}
    \label{fig:Finalfix}
  \end{subfigure}%
  \hfill
  \begin{subfigure}[t]{\linewidth}
    \centering
    \includegraphics[width=\linewidth]{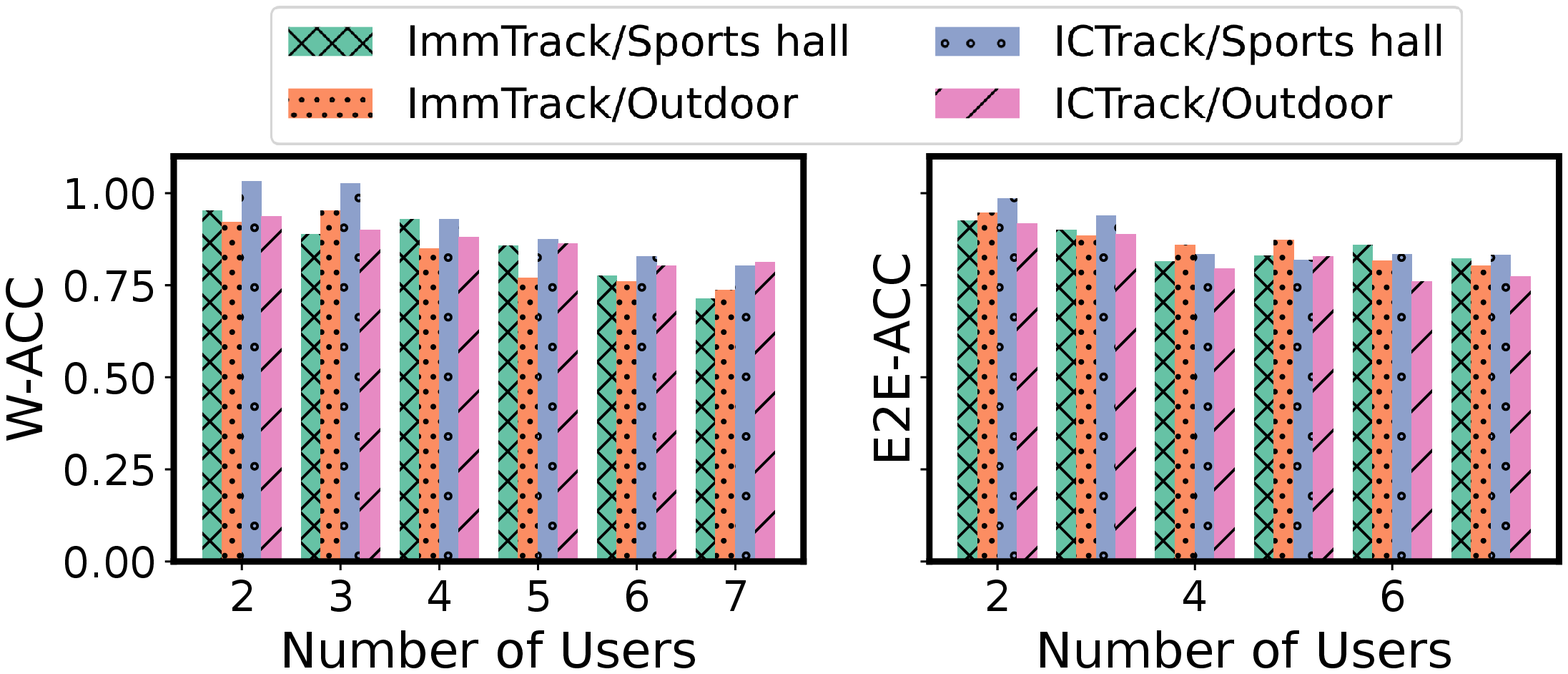}
    \caption{Sports hall and outdoor W-ACC and E2E-ACC}
    \label{fig:Finalfix2scene}
  \end{subfigure}%
  \vspace{-1em}
  \caption{Cross-modality association accuracy.}
  \label{fig:accFinal}
  \vspace{-1em}
\end{figure}

\subsubsection{Baselines and evaluation metrics}

We employ the following \yimin{three} baseline systems.


$\blacksquare$ {\em ICTrack} is the variant of ImmTrack with mmWave radar replaced by camera.
Camera provides much higher resolution than mmWave radar, but causes privacy concerns. 
ICTrack employs YOLO \cite{yolo} to detect objects and Deep SORT \cite{deepsort} to associate the bounding boxes of the same object in adjacent image frames.
In our implementation, the feature dimension used in Deep SORT for each bounding box is 416. However, Deep SORT does not exploit the prior information of the total number of users (i.e., $N$). As a result, it often mistakenly creates a new tracking identity for a previously seen user. For fair comparison, we explicitly correct a wrongly created tracking identity by the nearest bounding box in the previous frame.
ICTrack generates the 2D trajectory of each detected user from the video stream and executes the cross-modality trajectory association module presented in \sect\ref{subsec:association}.

\yimin{
  $\blacksquare$ {\em ImmTrack-ICP} is the variant of ImmTrack with the Siamese network replaced by {\em colored-ICP} \cite{park2017colored}, a colored point cloud registration algorithm. ImmTrack-ICP applies colored-ICP to find the optimal transformation matrix from each trace map of the radar cluster to each trace map of IMU. 
ImmTrack-ICP adopts the optimization objective function value of the transformation as the similarity between the trace maps of the radar cluster and IMU.
}

$\blacksquare$ {\em mmUniverSense} is a variant of UniverSense \cite{pan2018universense} that associates the user's limb movement detected by camera with IMU data based on movement acceleration.
We compare UniverSense's single metric-based association with ImmTrack's high-dimensional comparative feature-based association. For fair comparison, we adapt UniverSense to mmWave radar by replacing the acceleration metric with velocity metric, as mmWave radar directly provides velocity data.
This adapted version is called mmUniverSense.

{\em Evaluation metrics:} We adopt the ratio of correctly associated pairs to all users to characterize the association accuracy. This accuracy in each association time window is denoted by W-ACC, while the accuracy of the association achieved by the average similarity matrix over $W$ windows is called end-to-end accuracy (E2E-ACC).

\subsubsection{Association performance in sports hall and outdoor spaces} \label{main_acc_result}

\yimin{
Fig.~\ref{fig:Finalfix} presents W-ACC and E2E-ACC of ImmTrack, ImmTrack-ICP, ICTrack, and mmUniverSense on the data collected in the sports hall and outdoor spaces. For each setting of $N$, the experiment lasts for half an hour. Overall, ImmTrack achieves comparable performance with ICTrack on cross-modality association, while remaining less privacy-intrusive.
Specifically, ImmTrack achieves E2E-ACC from 81.4\% to 93.6\%, while ICTrack achieves 85.4\% to 95.1\%.
On W-ACC and E2E-ACC, ICTrack outperforms ImmTrack by around 7\% and 3\%, respectively.
The accuracy of mmUniverSense is inferior, because when users walk at similar speeds, the association merely based on velocity is prone to be erroneous. ImmTrack-ICP gives the lowest accuracy, which is close to random guessing. For each pair of trace maps from mmWave radar cluster and IMU, the colored-ICP algorithm  finds a transformation with small error even if the cluster and IMU are from different users. As a result, all values in the similarity matrix are high and the association process is close to random guessing.
}

\begin{figure}[t]
  \includegraphics[width=.85\linewidth]{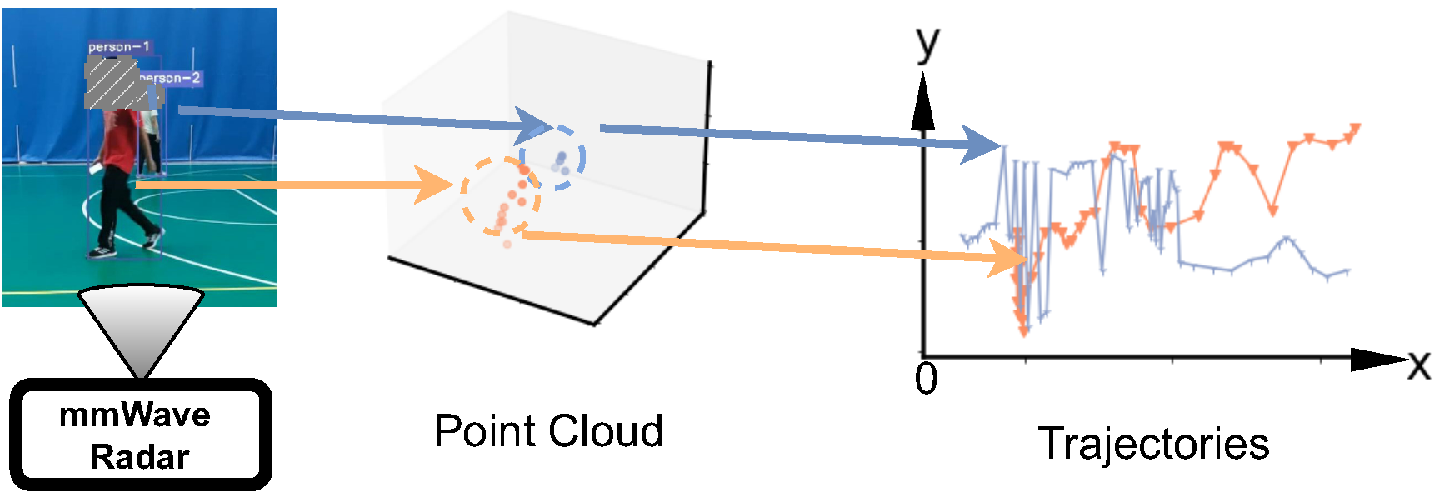}
  \vspace{-1em}
  \caption{ImmTrack can track the trajectory of a partially occluded user (marked in blue) correctly with help of IMU. } 
  \label{fig:occulsion-demo}
  \vspace{-1.5em}
\end{figure}

As shown in Fig.~\ref{fig:Finalfix2scene}, camera-based ICTrack yields higher accuracy indoors than outdoors. 
Essentially, the performance of ICTrack may degrade in certain environments with dimmed illumination, e.g., in museums with low illumination for protecting ancient artifacts. Differently, ImmTrack yields consistent accuracy, as mmWave radar is robust to different illumination condition.

By analyzing the results of ICTrack, 
YOLO in ICTrack performs well in detecting humans (as shown in Fig.~\ref{fig:subject-detection}), while Deep SORT has difficulties in associating bounding box across frames due to the non-coherent visual features of the same user in different frames. Differently, ImmTrack employs extensive features including shape, motion and IMU PID to achieve robust inter-frame cluster tracking. 
Note that the experiments include cases of inter-person occlusions.
In Fig.~\ref{fig:occulsion-demo}, we show that the mmWave radar can still yield some points on the visually occluded user, though with a lower density. This, together with our IMU-assisted design, makes ImmTrack work well in the transient occlusion cases.

\subsubsection{Dealing with passengers entering the monitored space} \label{eval_passenger}

  

\begin{figure}
    \centering
    \begin{subfigure}[t]{.46\linewidth}
    \centering
  \includegraphics[width=.8\linewidth]{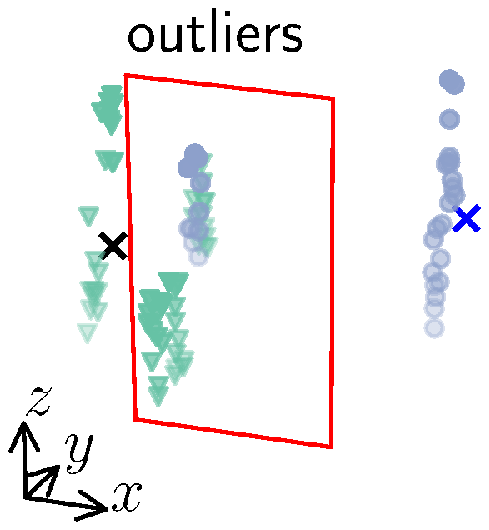}
  \caption{ImmTrack's clustering with passenger causing outliers}
  \label{fig:clusterP1}
  \end{subfigure}
  \hfill
  \begin{subfigure}[t]{.5\linewidth}
       \includegraphics[width=\textwidth]{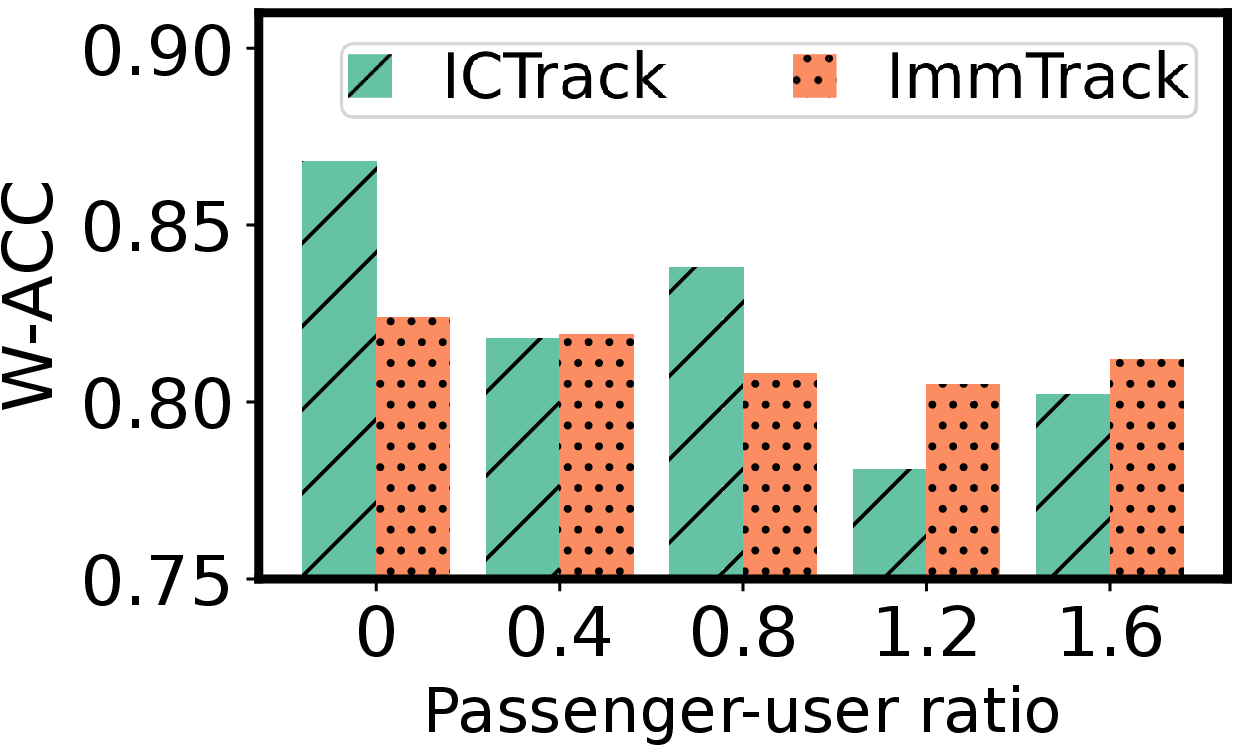}
    \caption{W-ACC vs. the ratio of passengers to users}   
    \label{fig:accEachWin5Pass}
  \end{subfigure}       
  
  \vspace{-1em}
\caption[ ]
{Impact of passengers on cross-modality association.}
\label{fig:accEachWinAllpass}
\vspace{-1em}
  \end{figure}

A passenger refers to a person who is within the monitored space but does not participate in the monitoring. 
For instance, a person whose smartphone is not installed with the ImmTrack app is a passenger. 
In the presence of passengers, there are outlier points corresponding to the passengers away from the new centroids after the RKF-assisted $k$-means clustering. 
To address this problem, ImmTrack views all the points out of the new centroids' bounding boxes as outliers and removes them, where the bounding box size is set to be commensurate to human body dimension. 
This design is motivated by the fact that the enhanced RKF-assisted $k$-means algorithm can keep tracking the users even if passengers enter the space, as long as ImmTrack is bootstrapped from a situation with no passenger. 
Fig.~\ref{fig:clusterP1} shows ImmTrack's clustering when one out of three people is a passenger. The outlier points away from the centroids represented by crosses are excluded from the clustering result.
For fair comparison, we also augment ICTrack to deal with passenger. In specific, we use an asymmetric auction algorithm to perform the $M$-to-$N$ bipartite cross-modality matching, where $M$ is the total number of people detected by YOLO, and $N$ is the number of users. 
We measure W-ACC when a certain number (0 to 8) of passengers enter the monitored space, while fixing the number of users at 5. 
From Fig.~\ref{fig:accEachWin5Pass},
ImmTrack achieves similar or even better W-ACC than ICTrack when there are passengers; the W-ACC of ImmTrack is not sensitive to the passenger-user ratio.

\begin{table}[]
\caption{Performance improvement by one more radar.}
  \label{tab:2radar-improve}
  \vspace{-1em}
  \begin{tabular}{ccccccc}
  \hline
  Number of users & 2     & 3     & 4     & 5     & 6     & 7     \\ \hline
  $\!\!$W-ACC improvement   & 2.3\% & 2.1\% & 4.0\% & 2.5\% & 4.2\% & 3.8\% \\
  $\!\!$E2E-ACC improvement$\!\!$ & 1.1\% & 1.2\% & 1.1\% & 1.0\% & 1.7\% & 1.3\% \\ \hline
  \end{tabular}
\end{table}

\subsubsection{Combining point clouds from multiple radars}\label{subsec:multi-radar}


Properly combining the point clouds from multiple radars may increase the spatial coverage of a space as well as the point density of a human target seen by multiple radars. 
In this set of experiments, we deploy two radars with their FOVs' axes of symmetry perpendicular. 
To accurately combine the two point clouds, we first apply a linear transform including a 90\textdegree{} rotation and origin shift to one point cloud, such that the two point clouds are roughly aligned. Then, we apply the iterative closest point (ICP) algorithm to perform a fine registration of the two point clouds. 
Table~\ref{tab:2radar-improve} presents the W-ACC and E2E-ACC improvement over varying number of users when two radars are used. With one more radar, there are about 4\% and 1\% absolute improvements in W-ACC and E2E-ACC, respectively, due to higher point cloud density.

\begin{figure}
 
  \begin{subfigure}[t]{0.5\linewidth}
    \centering
    \includegraphics[width=\linewidth]{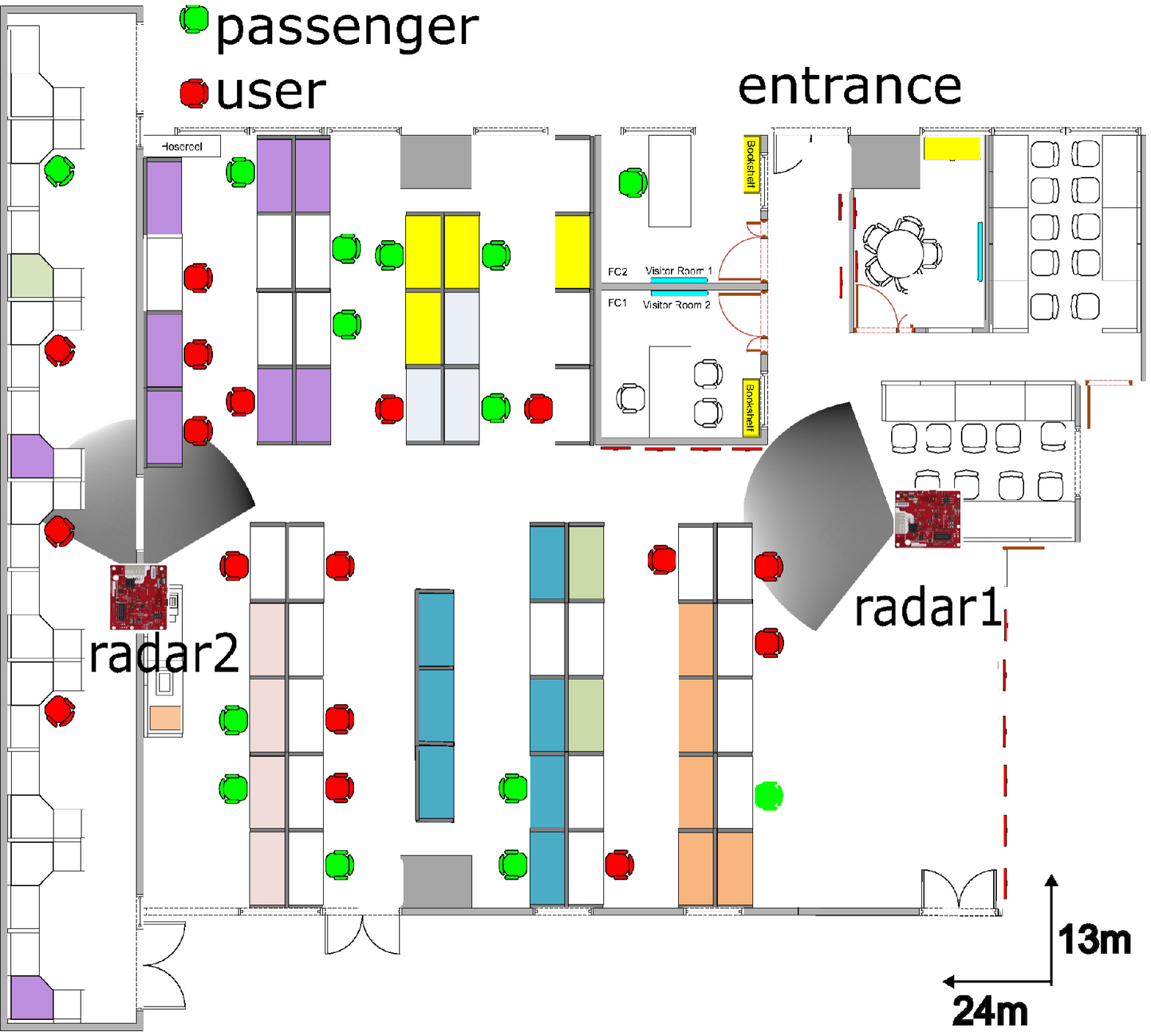}
   \caption{Floor plan; angular coverages of two radars; a snapshot of distribution of human subjects.}
    \label{Fig:floorPlan}
  \end{subfigure}%
  \hfill
  \begin{subfigure}[t]{.48\linewidth}
    \centering
    \includegraphics[width=\linewidth]{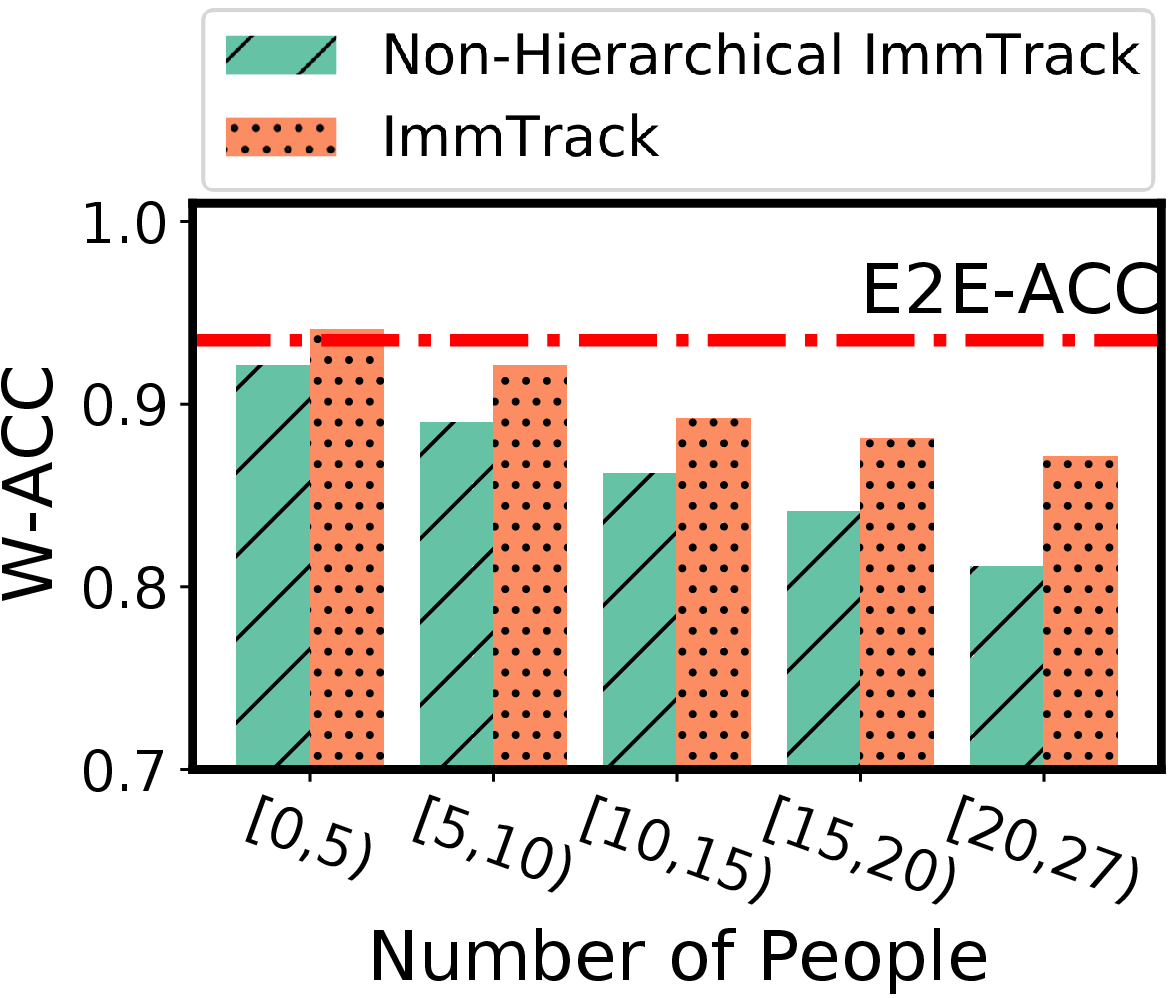}
   \caption{W-ACC vs. number of people. Horizontal line indicates mean E2E-ACC over four days. }
    \label{Fig:end2endcncl}
  \end{subfigure}%
  \vspace{-1em}
  \caption{Cross-modality association in a live lab space.}
  \label{fig:accTinyFinal}
\end{figure}

\subsubsection{Evaluation in a live lab space}
\label{eval_tiny_space}

Fig.~\ref{Fig:floorPlan} shows the floor plan. The total area of the space is about $300\,\text{m}^2$. We deploy two mmWave radars to fully cover the corridors and occupied workspaces, while accounting for the blockages caused by internal concrete structures. A total of 17 lab residents voluntarily participate in our evaluation by installing the IMU data collection program on their smartphones. Other lab residents are passengers to our system. \yimin{During the timespan, the numbers of users and passengers in the lab change.}
Fig.~\ref{Fig:floorPlan} also shows a snapshot distribution of the users and passengers. We collect data for four consecutive days. In this setup, we observe the users may walk side by side in the corridor. Thus, we particularly evaluate the effectiveness of the mechanism presented in \sect\ref{boundary_case_macth} for handling identical trace maps. The ImmTrack variant that does not apply the mechanism to separately process the nearly identical trace maps is called {\em non-hierarchical ImmTrack}. 
Note that stationary users, who can be detected in both the radar and IMU modalities, are excluded from the processing pipeline, because the workspaces in this lab conform to safe distancing requirement.
\yimin{However, the stationary users' locations and PIDs are maintained in the system.}
Fig.~\ref{Fig:end2endcncl} shows the W-ACC of ImmTrack and the non-hierarchical ImmTrack, versus the total number of people in the monitored area. \yimin{The $x$-axis is the number of people in the lab during different
testing periods.}
ImmTrack achieves up to 5.6\% higher W-ACC compared with the non-hierarchical ImmTrack. The horizontal line in Fig.~\ref{Fig:end2endcncl} shows the mean E2E-ACC of ImmTrack over the entire evaluation period, which is 94.1\%.
\subsection{Distance Tracking and Contact Tracing } \label{casestudy}


\yimin{We compare the interpersonal distance tracking performance of ImmTrack with the performance of mmTrack \cite{wu2020mmtrack}. In addition, we evaluate ImmTrack's performance for contact tracing. }
We collect a 47-minute trace with mmWave and camera data recorded, where seven users move in the sports hall shown in Fig.~\ref{fig:different scene}. We \yimin{apply ICTrack and manually rectify ICTrack's tracking identities to generate de-anonymized groundtruth trajectories of all the users}. In addition, we project the trajectories to the world coordinate system based on the camera's setup geometry and calculate the \yimin{interpersonal distance in the global coordinate system} as the {\em reference} to evaluate the accuracy of ImmTrack's interpersonal distance tracking and contact tracing results.

\begin{figure}
\begin{minipage}{0.23\textwidth}
  \centering 
  \includegraphics[width=\textwidth]{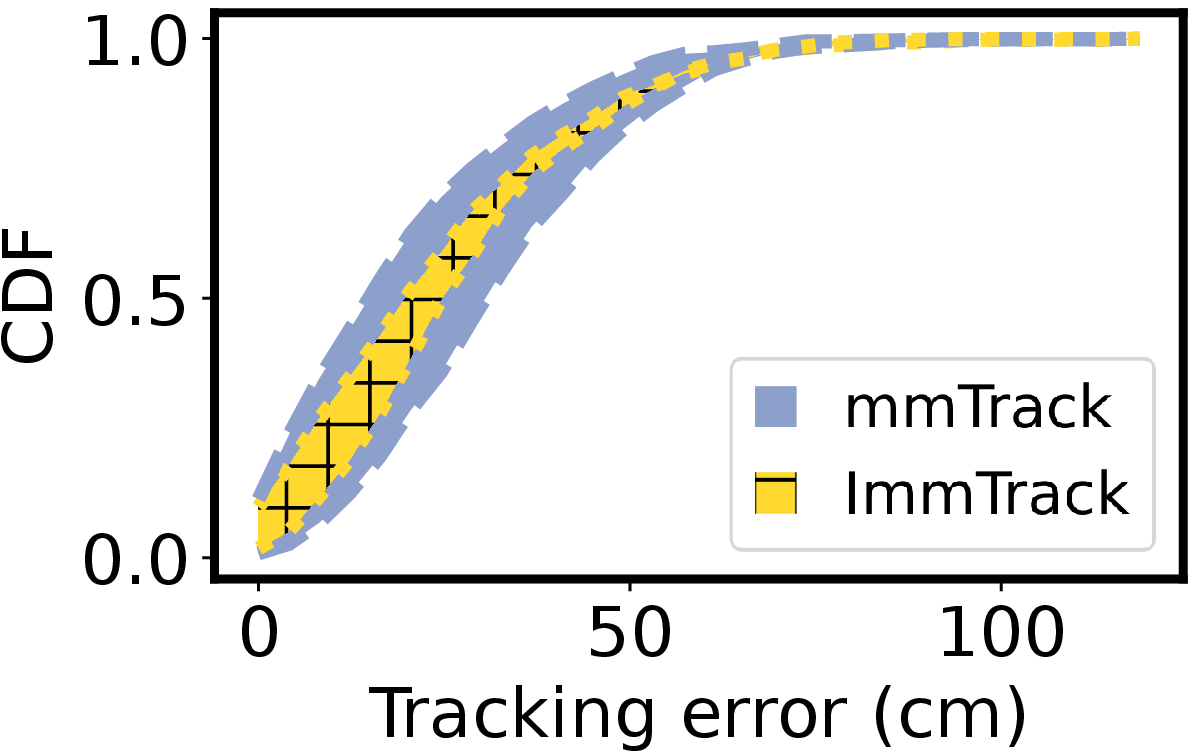}
  \vspace{-2 em}
  \caption[ ]{\yimin{Statistical analysis of ImmTrack's and mmTrack's tracking errors. The highlighted part of each color represents the area covered by the CDF curve of different users in a system.}}
  \label{fig:cdf_dis}
\end{minipage}%
\hfill
\begin{minipage}{0.23\textwidth}
\centering 
\includegraphics[width=\textwidth]{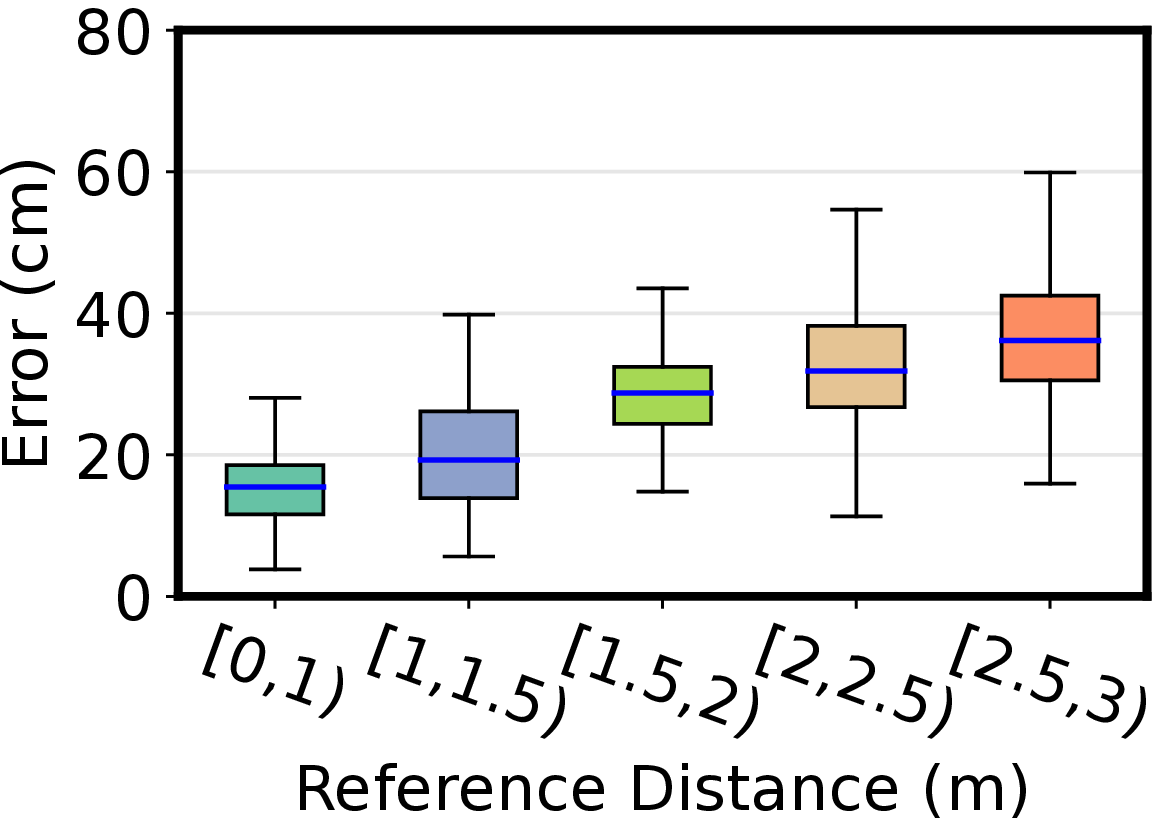}
\vspace{-1.6em}
\caption[ ]{ImmTrack's tracking error when reference distance is different. The horizontal line in the middle represents the average value of the error. }
\label{fig:contact_dis_diff}
\end{minipage}%
\vspace{-1em}
\end{figure}

$\blacksquare$ {\bf Spatial accuracy of interpersonal distance tracking.} \yimin{Fig.~\ref{fig:cdf_dis} shows the CDF of ImmTrack's and mmTrack's tracking errors in centimeters with respect to the reference trajectory.}
For ImmTrack, most tracking errors are within $50\,\text{cm}$. The average tracking error is $22\,\text{cm}$, showing that ImmTrack can achieve re-identified human tracking with decimeters spatial accuracy. \yimin{Compared with mmTrack, ImmTrack yields more stable tracking accuracy.}

For contact tracing, the tracking accuracy is important especially when the actual interpersonal distances are small. 
Fig.~\ref{fig:contact_dis_diff} shows ImmTrack's interpersonal distance tracking errors when the reference distance is in different ranges. When the reference distance is within one meter, the tracking errors are within $28\,\text{cm}$ and the mean error is $14\,\text{cm}$. The mean error remains under $40\,\text{cm}$ when the reference distance is up to $3\,\text{m}$. These results show that ImmTrack can accurately track interpersonal distances in close contacts.

\begin{figure*}
  \centering
  \begin{subfigure}[t]{0.3\textwidth}   
      \centering 
      \includegraphics[width=\textwidth]{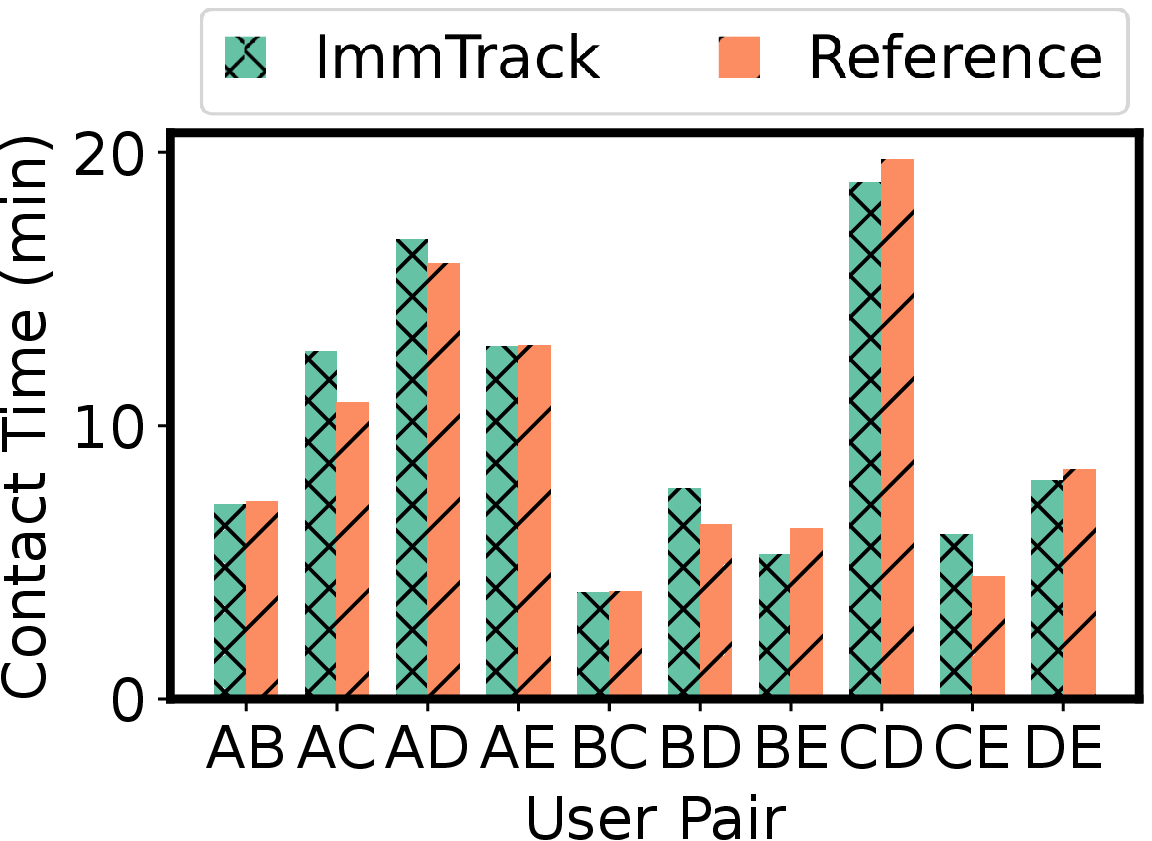}
      \caption[ ]%
    {Accumulative close contact time estimation between any two users in the 47-minute experiment $(N=5)$.}
    \label{fig:contact_time_compare}
  \end{subfigure}
  \hfill
  \begin{subfigure}[t]{.3\textwidth}
    \centering
    \includegraphics[width=\linewidth]{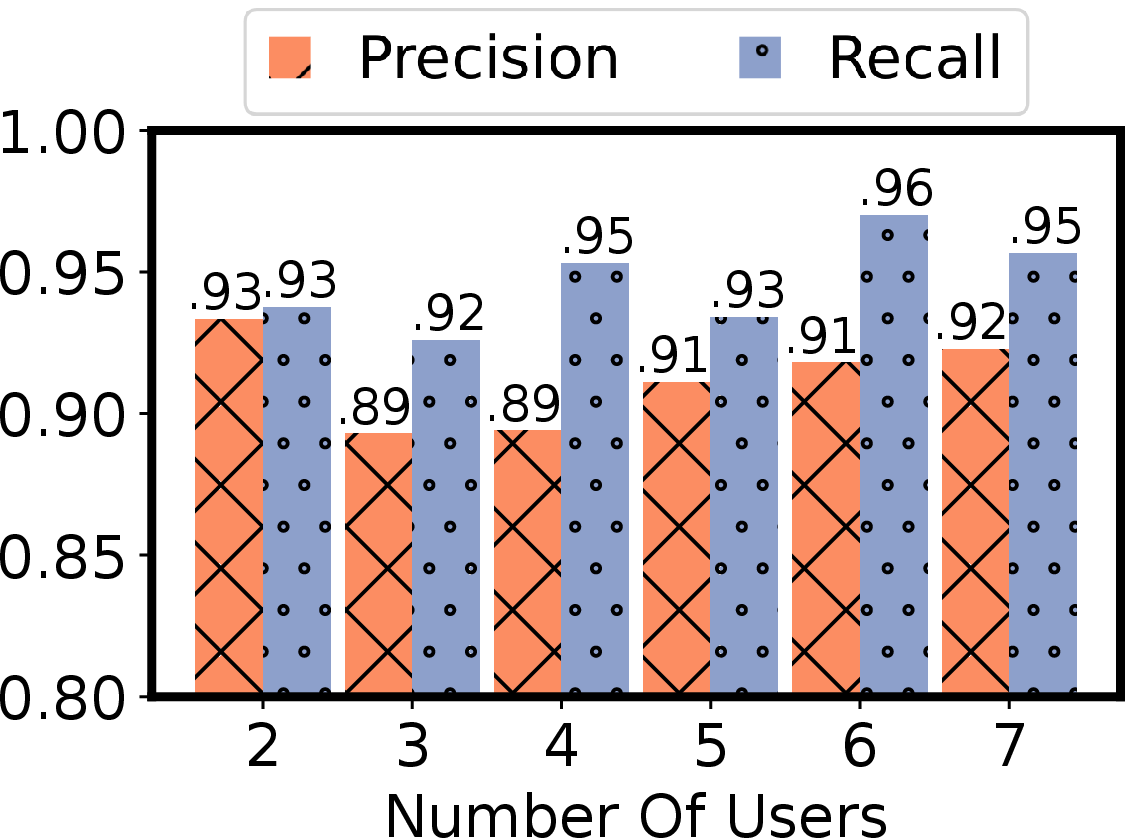}
    \caption[]{ImmTrack's precision and recall in pinpointing infectious contacts versus $N$ ($\tau=6\,\text{s}$).}
    \label{fig:contact_error}
  \end{subfigure}%
  \hfill
  \begin{subfigure}[t]{.3\textwidth}
    \centering
    \includegraphics[width=\linewidth]{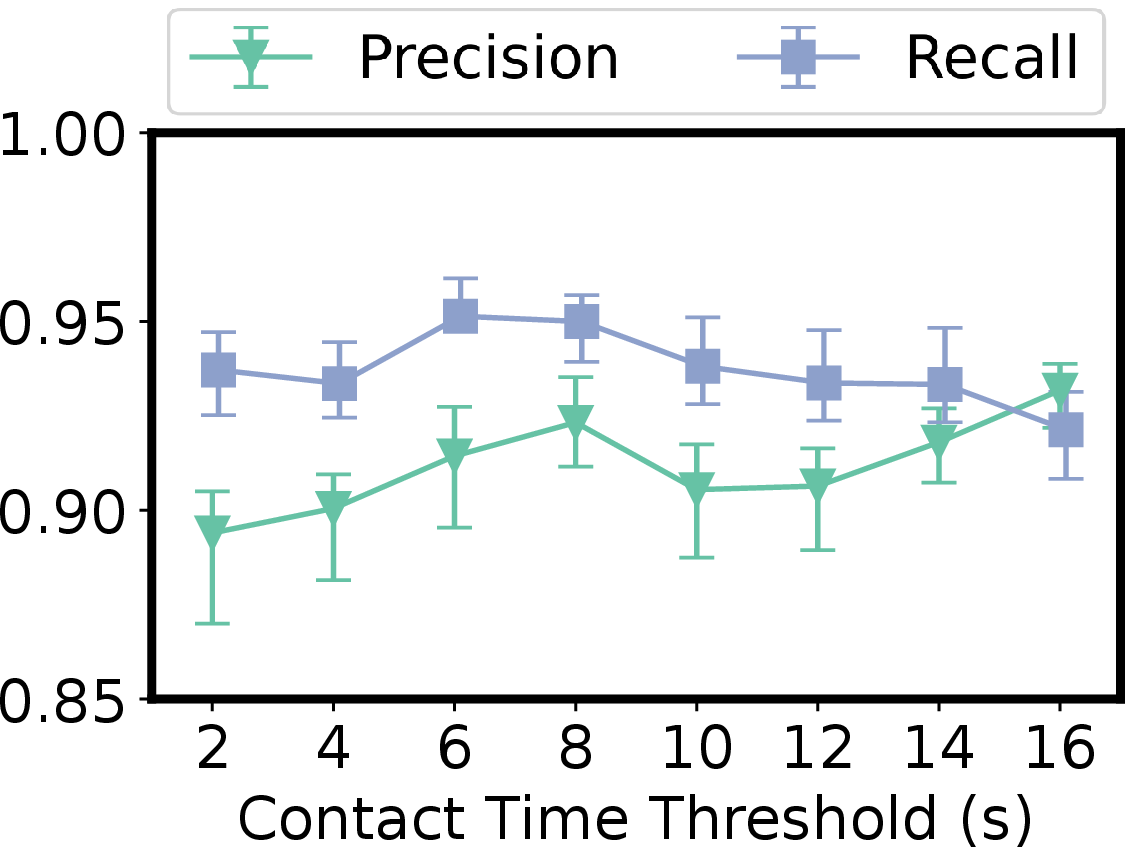}
    \caption[]{\yimin{ImmTrack's precision and recall in pinpointing infectious contacts under different contact time threshold $\tau$.}}
    \label{fig:contact_error_time}
  \end{subfigure}%
  \vspace{-1em}
  \caption[ ]
  {ImmTrack's performance on contact time estimation and pinpointing infectious contacts.}
  \label{fig:casestudy}
\end{figure*}


$\blacksquare$ {\bf Contact tracing performance.}
We consider two definitions of contact: (1) By following a prevailing definition, a {\em close contact} is a contact with less than $2\,\text{m}$ interpersonal distance; (2) An {\em infectious contact} is a contact with less than $1\,\text{m}$ interpersonal distance over $\tau$ seconds or more, where we set $\tau$ from 2 to \yimin{16} seconds.
Fig.~\ref{fig:contact_time_compare} shows the accumulative close contact time for each pair of users during the 47-minute experiment. It shows that ImmTrack's result and the reference. We can see that ImmTrack gives satisfactory close contact monitoring accuracy.
Then, we evaluate ImmTrack's performance in pinpointing infectious contact.
We slide a time window of $\tau+2$ seconds with two seconds overlapping and check whether an infectious contact occurs between any two users in the window.
By checking against the reference result in each time window, ImmTrack's detection result is among the true/false positive/negative.
We measure the {\em precision} and {\em recall} by $\text{precision}=\frac{\text{\# of true positives}}{\text{\# of all ImmTrack's positives}}$ and $\text{recall}=\frac{\text{\# of true positives}}{\text{\# of all reference's positives}}$. 
Fig.~\ref{fig:contact_error} shows the precision and recall for $\tau=6\,\text{s}$ when $N$ varies. Note that for each $N$ setting, we conduct a separate experiment that lasts for about 47 minutes. ImmTrack achieves about 90\% precision and \yimin{91\%-96\%} recall in pinpointing infectious contacts. \yimin{The opposite trend of recall and precision is due to the increase in the proportion of false negatives in all reference contacts.}


$\blacksquare$ {\bf Temporal resolution of contact tracing.} We vary the setting of $\tau$ to investigate the temporal resolution of ImmTrack in contact tracing.
Fig.~\ref{fig:contact_error_time} shows the precision and recall in pinpointing infectious contact versus the $\tau$ setting. While the recall remains stable at around 94\%, the precision increases from about 90\% to 93\% when $\tau$ is from 2 to \yimin{16} seconds. This shows that ImmTrack can achieve satisfactory temporal resolution fine to 2 seconds with a little contact detection accuracy drop. 
For comparison, we measure the BND detection delays using two or five Android phones. When using five phones, we place them at vertexes of a pentagon. Table~\ref{tab:bledelay} shows the time for a phone to discover all other phones versus the distance between the two phones or side length of the pentagon. The discovery delay increases with the distance and the number of phones. When the distance is one and three meters, the measured worst-case delay is more than 30 and 80 seconds, respectively.

\begin{table}[]
  \caption{BND detection delay (s) vs. inter-user distance (m).}
  \label{tab:bledelay}
  \vspace{-1em}
  \begin{tabular}{cccc}
  \hline
  Min inter-user distance & [0,1)          & [1,2)         & [2,3)          \\ \hline
  Two users                                                          & 3.2$\pm$2.1 & 4.8$\pm$1.7 & 6.9$\pm$3.7   \\
  Five users                                                         & 11.0$\pm$4.0 & 23.9$\pm$9.8 & 42.9$\pm$14.1 \\ \hline
  \end{tabular}
  \vspace{-1em}
  \end{table}


\subsection{Training and Efficacy of mmClusterNet}\label{compare_down_task}

  \begin{table}
  \caption{Summary of training datasets \& downstream tasks.}
  \label{tab:summary_ds_task}
  \vspace{-1em}
    \begin{tabular}{|c|c|c|c|}
      \hline
      \textbf{Model} & \textbf{Input} & \textbf{Training} & \textbf{Downstream} \\
                 &    & \textbf{dataset} & \textbf{task} \\ \hline
      \multirow{3}{*}{mmClusterNet} & \multirow{3}{*}{\begin{tabular}[c]{@{}c@{}}Point cloud\\with velocity\end{tabular}} & \multirow{3}{*}{\begin{tabular}[c]{@{}c@{}}Self-\\collected\end{tabular}} & PC                                                                               \\ \cline{4-4} 
                     &  &                                                                          & BBR                \\ \cline{4-4}
                     &  &                                                                          & NBBR                                                                                   \\ \hline
      \multirow{2}{*}{PointNet}                                              &  \multirow{2}{*}{\begin{tabular}[c]{@{}c@{}}Point cloud\\w/o velocity\end{tabular}} & \multirow{2}{*}{ShapeNet}                                                  & OC                                    \\ \cline{4-4} 
                     &    &                                                                        & PC                                                                           \\ \hline      
    \end{tabular}
\end{table}

The MLPs used by mmClusterNet to extract the shape-motion feature of a point cloud cluster needs to be trained before use. The training requires a downstream task that utilizes the shape-motion feature. This set of experiments evaluates the impact of various downstream tasks on the training of mmClusterNet. We also compare the cluster tracking feature extracted by mmClusterNet and the feature extracted by PointNet \cite{pointnet}, a widely adopted point cloud feature extractor. PointNet takes a point cloud without velocity as input and also needs a downstream task to drive training.

Table~\ref{tab:summary_ds_task} summarizes the input data, training datasets, and downstream tasks used to train mmClusterNet and PointNet. Beside the widely adopted point cloud completion (PC), bounding box regression (BBR), and object classification (OC) tasks, we devise a new task called {\em next-frame bounding box regression} (NBBR), which predicts the 2D bounding box with orientation in the next frame based on the feature extracted from the current frame. The loss functions used by the downstream tasks are as follows: PC uses chamber distance \cite{fan2017point}; BBR and NBBR use intersection over union (IoU); OC uses negative log likelihood.
We employ the multiple object tracking error (MOTE) and ratio of mismatches (RoM)
to jointly measure the inter-frame cluster tracking performance.
A mismatch refers to the case that a cluster is associated with another cluster in the previous frame that corresponds to a different user. 


\begin{figure}
  \begin{minipage}{0.21\textwidth}
  \centering
  \includegraphics[width=\linewidth]{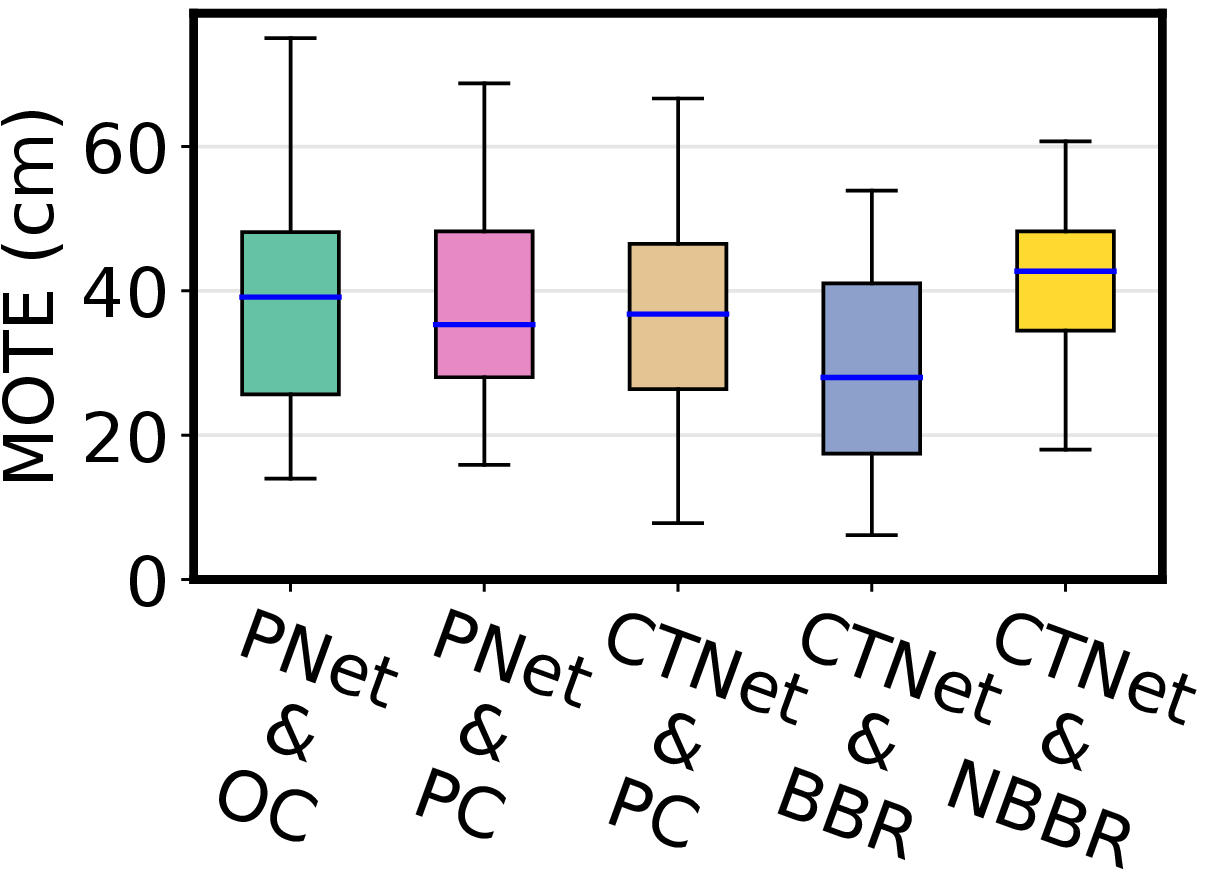}
  \vspace{-2em}
  \caption{Multi-object tracking error of inter-frame cluster tracking.}
  \label{fig:motp}
\end{minipage}%
\hfill
\begin{minipage}{0.23\textwidth}
  \centering
  \includegraphics[width=\linewidth]{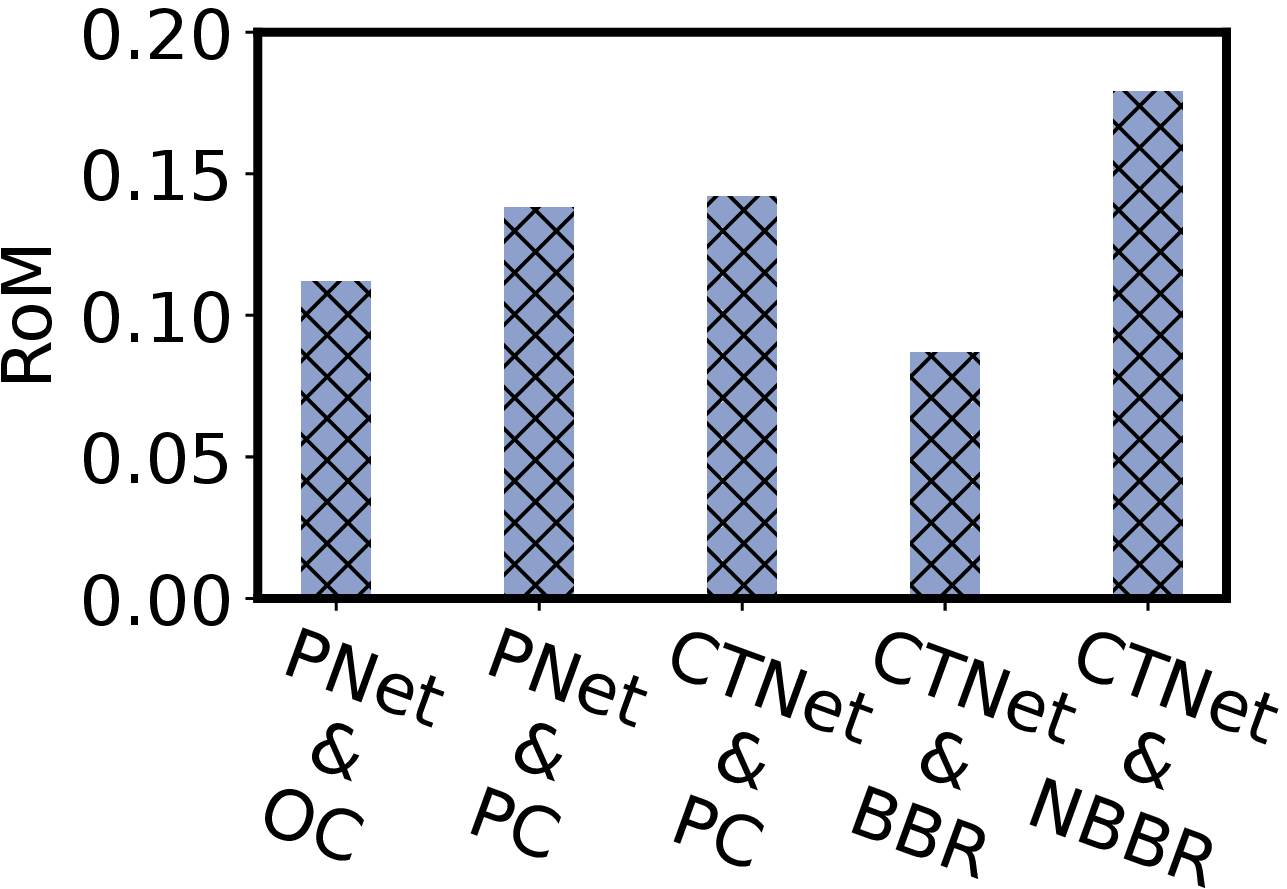}
  \vspace{-2em}
  \caption{Ratio of mismatches during inter-frame cluster tracking.}
  \label{fig:mismatch}
\end{minipage}%
\vspace{-1em}
\end{figure}


The results in Figs.~\ref{fig:motp} and \ref{fig:mismatch} show that: (1) mmClusterNet outperforms the off-the-shelf PointNet in achieving inter-cluster tracking; (2) BBR is an appropriate downstream task for training mmClusterNet.
BBR enforces the model to simultaneously capture cluster contour and enforces utilization of the velocity information of the shape-motion feature. Thus, BBR helps mmClusterNet better learn the shape-motion feature. On the contrary, NBBR leads to poor tracking performance. A possible reason is that NBBR overstretches the utilization of velocity information. ImmTrack evaluated in other sections adopts the mmClusterNet trained with BBR.

\subsection{Compute and Communication Overheads}
\label{subsec:time}

\begin{figure}
  \centering

  \begin{minipage}{.45\linewidth}%
    \includegraphics[width=\linewidth]{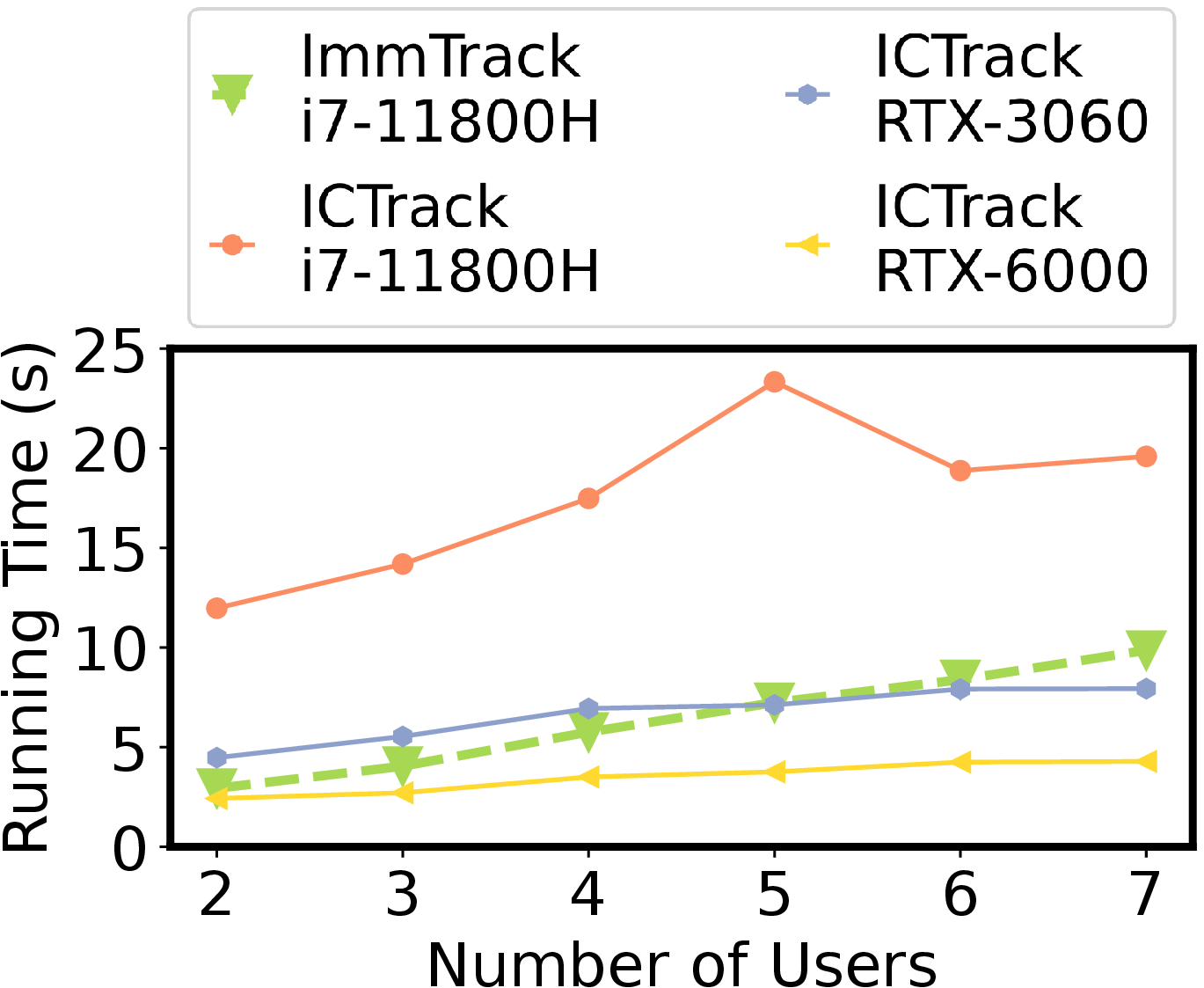}%
    \vspace{-1em}
    \caption{Runtime latency of ImmTrack and ICTrack with different hardwares.}
    \label{fig:timeFinal}%
  \end{minipage}
  \hfill
  \begin{minipage}{.52\linewidth}  
  \centering
    \includegraphics[width=.865\linewidth]{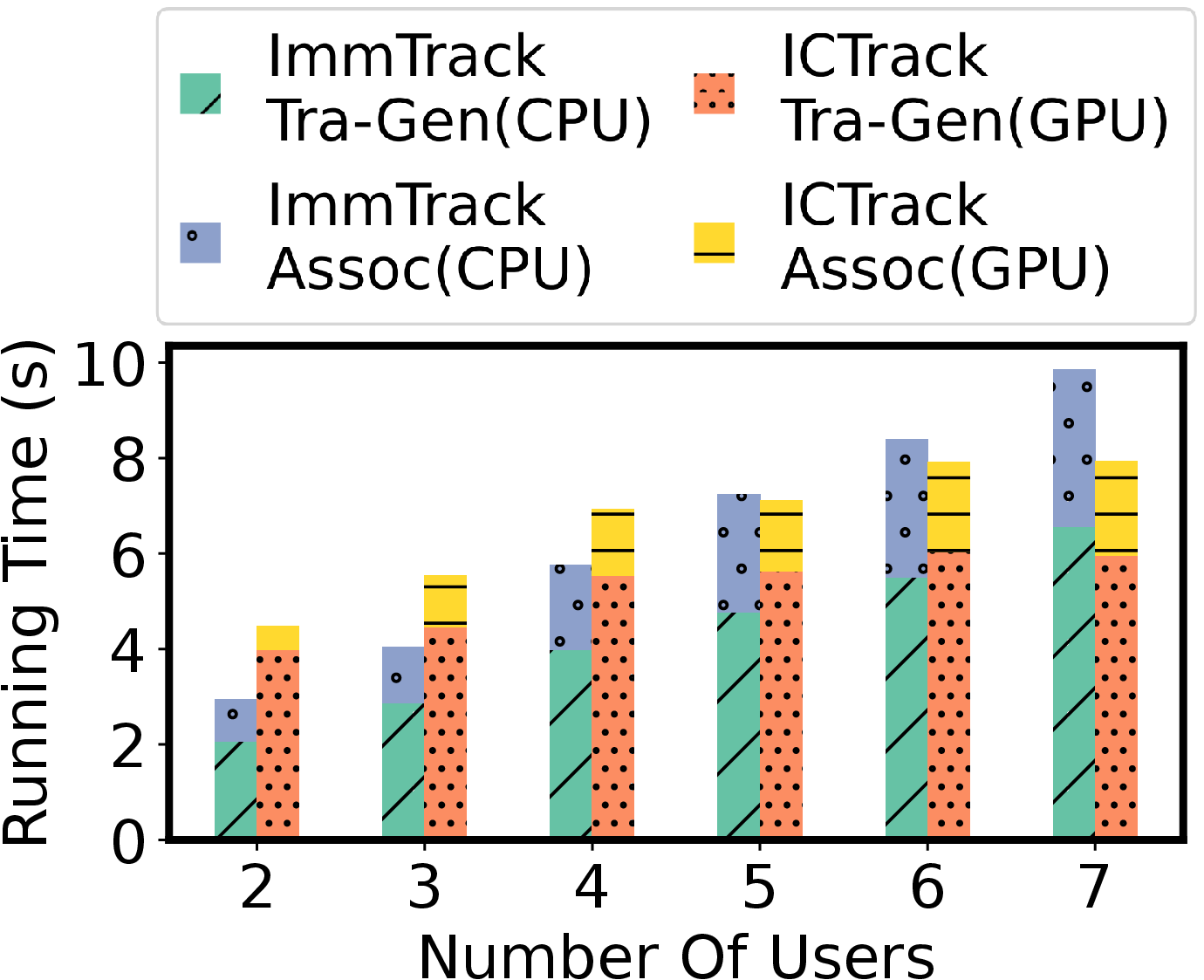}%
    \vspace{-1.2em}
    \caption{Time for trace map generation (Tra-Gen) and cross-modality association (Assoc).}
    \label{fig:timeDivide}%
  \end{minipage}%
  \vspace{-1em}
\end{figure}

\subsubsection{Server computation overhead}
Fig.~\ref{fig:timeFinal} shows the runtime latency of ImmTrack and ICTrack on the server under different $N$. In general, ImmTrack runnning on an Intel i7-11800H CPU can achieve 30 to 60 fps, depending on the number of users. Note that our ImmTrack implementation adopts a radar sampling rate of 8 fps. Thus, a CPU-only cloud server can support several ImmTrack tasks for different venues, or a CPU-only {\em in situ} edge server can support a single ImmTrack instance. ICTrack on the same i7-11800H CPU can only achieve about 15 fps processing throughput. Even with a GeForce RTX-3060 or RTX-6000 GPU, ICTrack's processing throughput is still lower than ImmTrack's, because the image processing imposes higher computation overhead than point cloud processing. By jointly considering the accuracy results obtained in \sect\ref{system_performance}, compared with ICTrack, ImmTrack achieves similar accuracy but only requires 1/4 to 1/2 processing power.
Fig.~\ref{fig:timeDivide} shows the breakdown of the time for processing 90 frames to generate trace map and perform cross-modality association, where generating trace map from radar and camera data takes most of the time.

\subsubsection{Smartphone communication and energy overheads}

We deploy both the IMU sampling and trace map generation modules on an Android smartphone and measure the overheads.
ImmTrack uploads the velocity magnitude to the server for the mmWave-IMU pre-matching. At the end of each association time window, ImmTrack uploads the trace map to the server, which is about $30\,\text{KB}$. The mmUniverSense uploads the 3D velocity continuously.
Our measurements show that ImmTrack's and mmUniverSense's bit rates are $7.36\,\text{kbps}$ and $15.63\,\text{kbps}$, respectively. ImmTrack's bit rate is lower than the $8\,\text{kbps}$ of G.729, an ITU's voice codec for bandwidth-constrained scenarios.

We also compare the battery energy usages of ImmTrack and three existing contact tracing mobile apps, i.e., TraceTogether, LeaveHomeSafe, Coronalert. We run these  apps in the background on an Android smartphone for eight hours. We factory-reset the smartphone before each benchmark. ImmTrack keeps sampling IMU, computing trace maps, and uploading data. From publicly available information, Coronalert (which is based on Google/Apple Exposure Notification system) and TraceTogether exchange Bluetooth messages with nearby devices; LeaveHomeSafe is a passive tracing tool based on QR code scanning. During each 8-hour benchmark, we use the tested app to scan valid QR codes every hour to mimic normal daily usages. 
According to our measurements,  battery energy usages of TraceTogether, LeaveHomeSafe, Coronalert are 55.62, 157.04, 37.37 mAh, respectively, while ImmTrack consumes 36.05 mAh.
Thus, ImmTrack imposes similar/lower battery energy overhead compared with the existing contact tracing apps.

\section{Conclusion}
\label{sec:conclude}

This paper presents ImmTrack, an interpersonal distance tracking system using one or more low-cost mmWave radar(s) and the IMUs of the users' smartphones. By associating the users' trajectories reconstructed from the mmWave radar and IMU sensing in terms of the trajectory features extracted by a Siamese neural network,
ImmTrack transfers the users' pseudo identities tagged to the IMU data to the radar's global-view sensing results.
Extensive experiments with up to 27 people show that ImmTrack achieves similar tracking accuracy and lower computation overhead compared with the more privacy-intrusive camera surveillance. ImmTrack achieves decimeters-seconds spatio-temporal accuracy in tracing contacts, outperforming the prevailing Bluetooth neighbor discovery approach that suffers inaccurate distance estimation and up to 80 seconds discovery delays in our experiments.





\begin{acks}
  This research is supported in part by the Ministry of Education, Singapore, under its Academic Research Fund Tier 1 (RG88/22), in part by the Innovation and Technology Commission of Hong Kong under Grant No.~GHP/126/19SZ, and in part by the Research Grants Council (RGC) of Hong Kong under Grant No.~14209619.
\end{acks}

\ignore{
\newpage
\LARGE \noindent \textbf{APPENDIX} \normalsize

More specifically, a JSON file is needed in Brick to initiate points and their tag, with the 2-tuple format \{ Brick Point, (Corresponding) RDF Tag \}, e.g., \{ brick: High\_AirSpeed\_Sensor, rdf: type owl: Class \} and \{ brick: High\_AirSpeed\_Sensor, rdfs: subClassOf brick: AirSpeed\_Sensor \} for the air-speed sensor. We leverage another JSON file to store the initial mapping from the point in the dataset to the point in Brick with another 2-tuple format \{ Dataset Point, Brick Point \}, e.g., \{ vel\_h,  High\_AirSpeed\_Sensor \}. Then we connect the two JSON files using the following code to manually update the graph of entities: ``graph.add( Dataset Point, RDF Tag, Brick Point)'', where the three parameters refer to the triple \{ subject, predicate, object \} in the graph of entries of Brick.

\begin{figure}[t]
\centering
\includegraphics[angle=0, width=0.4\textwidth]{figures/ValueVsError.pdf}
\vspace{-0.5cm}
\caption{Error rate as a function of the number of distinct values of the task attribute, where the number of maximum layers is set as two to avoid point explosion in the figure.}
\label{fig:ValueVsError}
\end{figure}

\textbf{Metadata Vs. Data.} A key different between metadata and data lies in the number of distinct values. As the descriptor of data, metadata has much less number of distinct values than data. To evaluate the impact of metadata, we evaluate task definition with different attributes which have varying numbers of distinct values. We set the maximum number of layers in the forest as two in order to avoid task explosion mentioned above and keep a lower running time. We show the error rate of using these task attributes in Fig.~\ref{fig:ValueVsError}. We also compute a trend line over the scatter graph. A linear line fits well with $R^2 = 0.795$, which is quite close to 1. That indicates a strong linear relationship between the error rate and the number of distinct value. In general, the error rate increases as the number of distinct values raises, which indicates that metadata (with less number of values) plays a more important role in task definition for multi-task thermal-comfort prediction. 

\begin{table}[t]
\caption{Different numbers of maximum layers in the forest: improvement over the previous layer and training time.}
\label{tab:maxLayer}
\begin{tabular}{|c|c|c|}
\hline
 Max. \# Layers & IMP over Previous Layer (\%) & Training Time (h) \\ \hline
1              & -                         & 0.27              \\ \hline
2              & 31\%                      & 3.86              \\ \hline
3              & 11\%                      & 8.37              \\ \hline
4              & -36\%*                    & 14.27*            \\ \hline
\end{tabular}
\begin{flushleft} \small
*Memory error occurs in this layer and the program stopped in the middle. We provide the best result found before error occurs.
\end{flushleft}
\end{table}

\textbf{Impact of Maximum Number of Layers.} An important factor of DUET is the maximum number of layers allowed in the forest. In Table~\ref{tab:maxLayer}, we show the minimum error rate and training time as the function of the maximum number of layers in the forest. First, we see that the training time increases significantly as the maximum number of layers increases and it can even lead to memory error when the number reaches four. That is because when the maximum number of layer increases, the number of separated datasets increases exponentially, which leads to much more tries of combining and training. Second, we see that when the maximum number of layers is larger than two, the improvement of error rate drops significantly and may even lead to a higher error rate due to a memory error. The top layers of attributes have a much greater impact than those in the bottom layers. It is wiser to leverage a 2/3-layer forest to achieve a lower error rate and at the same time less computing resources.

\begin{figure}[t]
\centering
\includegraphics[angle=0, width=0.4\textwidth]{figures/metadegreeVsError.pdf}
\vspace{-1cm}
\caption{Error rate as a function of meta degree of the task attribute, where the maximum number of layers is two.}
\label{fig:metadegreeVsError}
\end{figure}

\textbf{Impact of Minimum Meta Degree.} Another important factor of DUET is the number of minimum meta degree $\Lambda$ which is used to define metadata. According to our definition, merely those data item with meta degree higher than $\Lambda$ is recognized as metadata and used to extract the meta attribute and the final task attribute. In Fig.~\ref{fig:metadegreeVsError}, we plot the error rate as a function of meta degree. We see that basically, the error rate is inversely proportional to the meta degree. In RP884, with the maximum number of layers set as two, the best error rate can be achieved with the data item of VELAV, which has a meta degree as high as 0.9996.
}

\bibliographystyle{ACM-Reference-Format}
\bibliography{reference} 


\begin{thebibliography}{49}


\ifx \showCODEN    \undefined \def \showCODEN     #1{\unskip}     \fi
\ifx \showDOI      \undefined \def \showDOI       #1{#1}\fi
\ifx \showISBNx    \undefined \def \showISBNx     #1{\unskip}     \fi
\ifx \showISBNxiii \undefined \def \showISBNxiii  #1{\unskip}     \fi
\ifx \showISSN     \undefined \def \showISSN      #1{\unskip}     \fi
\ifx \showLCCN     \undefined \def \showLCCN      #1{\unskip}     \fi
\ifx \shownote     \undefined \def \shownote      #1{#1}          \fi
\ifx \showarticletitle \undefined \def \showarticletitle #1{#1}   \fi
\ifx \showURL      \undefined \def \showURL       {\relax}        \fi
\providecommand\bibfield[2]{#2}
\providecommand\bibinfo[2]{#2}
\providecommand\natexlab[1]{#1}
\providecommand\showeprint[2][]{arXiv:#2}

\bibitem[GAE(2022)]%
        {GAEN}
 \bibinfo{year}{2022}\natexlab{}.
\newblock \bibinfo{title}{Google/Apple Exposure Notifications}.
\newblock
\newblock
\newblock
\shownote{\url{https://www.google.com/covid19/exposurenotifications/}}.


\bibitem[Ahuja et~al\mbox{.}(2021)]%
        {vid2doppler}
\bibfield{author}{\bibinfo{person}{Karan Ahuja}, \bibinfo{person}{Yue Jiang},
  \bibinfo{person}{Mayank Goel}, {and} \bibinfo{person}{Chris Harrison}.}
  \bibinfo{year}{2021}\natexlab{}.
\newblock \showarticletitle{Vid2Doppler: Synthesizing Doppler Radar Data from
  Videos for Training Privacy-Preserving Activity Recognition}. In
  \bibinfo{booktitle}{\emph{CHI}}.
\newblock


\bibitem[Alahi et~al\mbox{.}(2015)]%
        {rgbw}
\bibfield{author}{\bibinfo{person}{Alexandre Alahi}, \bibinfo{person}{Albert
  Haque}, {and} \bibinfo{person}{Li Fei-Fei}.} \bibinfo{year}{2015}\natexlab{}.
\newblock \showarticletitle{RGB-W: When Vision Meets Wireless}. In
  \bibinfo{booktitle}{\emph{ICCV}}.
\newblock


\bibitem[Basso et~al\mbox{.}(2017)]%
        {basso2017kalman}
\bibfield{author}{\bibinfo{person}{Gabriel~F Basso}, \bibinfo{person}{Thulio
  Guilherme~Silva De~Amorim}, \bibinfo{person}{Alisson~V Brito}, {and}
  \bibinfo{person}{Tiago~P Nascimento}.} \bibinfo{year}{2017}\natexlab{}.
\newblock \showarticletitle{Kalman filter with dynamical setting of optimal
  process noise covariance}.
\newblock \bibinfo{journal}{\emph{IEEE Access}}  \bibinfo{volume}{5}
  (\bibinfo{year}{2017}), \bibinfo{pages}{8385--8393}.
\newblock


\bibitem[Boulic et~al\mbox{.}(1990)]%
        {boulic1990global}
\bibfield{author}{\bibinfo{person}{Ronan Boulic},
  \bibinfo{person}{Nadia~Magnenat Thalmann}, {and} \bibinfo{person}{Daniel
  Thalmann}.} \bibinfo{year}{1990}\natexlab{}.
\newblock \showarticletitle{A global human walking model with real-time
  kinematic personification}.
\newblock \bibinfo{journal}{\emph{Vis Comput}} \bibinfo{volume}{6},
  \bibinfo{number}{6} (\bibinfo{year}{1990}), \bibinfo{pages}{344--358}.
\newblock


\bibitem[Chiu et~al\mbox{.}(2020)]%
        {chiu2020impact}
\bibfield{author}{\bibinfo{person}{N.-C. Chiu}, \bibinfo{person}{H. Chi},
  \bibinfo{person}{Y.-L. Tai}, \bibinfo{person}{C.-C. Peng}, {et~al\mbox{.}}}
  \bibinfo{year}{2020}\natexlab{}.
\newblock \showarticletitle{Impact of wearing masks, hand hygiene, and social
  distancing on influenza, enterovirus, and all-cause pneumonia during the
  coronavirus pandemic: retrospective national epidemiological surveillance
  study}.
\newblock \bibinfo{journal}{\emph{J. Medical Internet Research}}
  \bibinfo{volume}{22}, \bibinfo{number}{8} (\bibinfo{year}{2020}),
  \bibinfo{pages}{e21257}.
\newblock


\bibitem[Fan et~al\mbox{.}(2017)]%
        {fan2017point}
\bibfield{author}{\bibinfo{person}{Haoqiang Fan}, \bibinfo{person}{Hao Su},
  {and} \bibinfo{person}{Leonidas~J Guibas}.} \bibinfo{year}{2017}\natexlab{}.
\newblock \showarticletitle{A point set generation network for 3d object
  reconstruction from a single image}. In \bibinfo{booktitle}{\emph{CVPR}}.
\newblock


\bibitem[Fang et~al\mbox{.}(2020)]%
        {eyefi}
\bibfield{author}{\bibinfo{person}{S. Fang}, \bibinfo{person}{T. Islam},
  \bibinfo{person}{S. Munir}, {and} \bibinfo{person}{S. Nirjon}.}
  \bibinfo{year}{2020}\natexlab{}.
\newblock \showarticletitle{EyeFi: Fast Human Identification Through Vision and
  WiFi-based Trajectory Matching}. In \bibinfo{booktitle}{\emph{DCOSS}}.
\newblock


\bibitem[Feng et~al\mbox{.}(2014)]%
        {feng2014kalman}
\bibfield{author}{\bibinfo{person}{Bo Feng}, \bibinfo{person}{Mengyin Fu},
  \bibinfo{person}{Hongbin Ma}, \bibinfo{person}{Yuanqing Xia}, {and}
  \bibinfo{person}{Bo Wang}.} \bibinfo{year}{2014}\natexlab{}.
\newblock \showarticletitle{Kalman filter with recursive covariance
  estimation—Sequentially estimating process noise covariance}.
\newblock \bibinfo{journal}{\emph{IEEE Trans. Ind. Electron.}}
  \bibinfo{volume}{61}, \bibinfo{number}{11} (\bibinfo{year}{2014}),
  \bibinfo{pages}{6253--6263}.
\newblock


\bibitem[Han et~al\mbox{.}(2016)]%
        {han2016amil}
\bibfield{author}{\bibinfo{person}{Hao Han}, \bibinfo{person}{Shanhe Yi},
  \bibinfo{person}{Qun Li}, \bibinfo{person}{Guobin Shen},
  \bibinfo{person}{Yunxin Liu}, {and} \bibinfo{person}{Ed Novak}.}
  \bibinfo{year}{2016}\natexlab{}.
\newblock \showarticletitle{AMIL: Localizing neighboring mobile devices through
  a simple gesture}. In \bibinfo{booktitle}{\emph{INFOCOM}}.
\newblock


\bibitem[He et~al\mbox{.}(2015)]%
        {he2015full}
\bibfield{author}{\bibinfo{person}{Shibo He}, \bibinfo{person}{Dong-Hoon Shin},
  \bibinfo{person}{Junshan Zhang}, \bibinfo{person}{Jiming Chen}, {and}
  \bibinfo{person}{Youxian Sun}.} \bibinfo{year}{2015}\natexlab{}.
\newblock \showarticletitle{Full-view area coverage in camera sensor networks:
  Dimension reduction and near-optimal solutions}.
\newblock \bibinfo{journal}{\emph{IEEE Trans. Veh. Technol.}}
  \bibinfo{volume}{65}, \bibinfo{number}{9} (\bibinfo{year}{2015}),
  \bibinfo{pages}{7448--7461}.
\newblock


\bibitem[He and Shin(2017)]%
        {he2017geomagnetism}
\bibfield{author}{\bibinfo{person}{S. He} {and} \bibinfo{person}{K. Shin}.}
  \bibinfo{year}{2017}\natexlab{}.
\newblock \showarticletitle{Geomagnetism for smartphone-based indoor
  localization: Challenges, advances, and comparisons}.
\newblock \bibinfo{journal}{\emph{ACM Comput. Surv}} \bibinfo{volume}{50},
  \bibinfo{number}{6} (\bibinfo{year}{2017}), \bibinfo{pages}{1--37}.
\newblock


\bibitem[Huang et~al\mbox{.}(2013)]%
        {huang2013connected}
\bibfield{author}{\bibinfo{person}{Hua Huang}, \bibinfo{person}{Chien-Chun Ni},
  \bibinfo{person}{Xiaomeng Ban}, \bibinfo{person}{Jie Gao}, {and}
  \bibinfo{person}{Shan Lin}.} \bibinfo{year}{2013}\natexlab{}.
\newblock \showarticletitle{Connected wireless camera network deployment with
  visibility coverage}. In \bibinfo{booktitle}{\emph{IPSN}}.
\newblock


\bibitem[Huang et~al\mbox{.}(2021)]%
        {huang2021indoor}
\bibfield{author}{\bibinfo{person}{Xu Huang}, \bibinfo{person}{Hasnain Cheena},
  \bibinfo{person}{Abin Thomas}, {and} \bibinfo{person}{Joseph~KP Tsoi}.}
  \bibinfo{year}{2021}\natexlab{}.
\newblock \showarticletitle{Indoor Detection and Tracking of People Using
  mmWave Sensor}.
\newblock \bibinfo{journal}{\emph{J. Sensors}} (\bibinfo{year}{2021}).
\newblock


\bibitem[Kempfle and Van~Laerhoven(2021)]%
        {kempfle2021quaterni}
\bibfield{author}{\bibinfo{person}{J. Kempfle} {and} \bibinfo{person}{K.
  Van~Laerhoven}.} \bibinfo{year}{2021}\natexlab{}.
\newblock \showarticletitle{Quaterni-On: Calibration-free Matching of Wearable
  IMU Data to Joint Estimates of Ambient Cameras}. In
  \bibinfo{booktitle}{\emph{PerCom}}.
\newblock


\bibitem[Kindt et~al\mbox{.}(2021)]%
        {kindt2021reliable}
\bibfield{author}{\bibinfo{person}{Philipp~H Kindt}, \bibinfo{person}{Trinad
  Chakraborty}, {and} \bibinfo{person}{Samarjit Chakraborty}.}
  \bibinfo{year}{2021}\natexlab{}.
\newblock \showarticletitle{How reliable is smartphone-based electronic contact
  tracing for COVID-19?}
\newblock \bibinfo{journal}{\emph{Commun. ACM}} \bibinfo{volume}{65},
  \bibinfo{number}{1} (\bibinfo{year}{2021}), \bibinfo{pages}{56--67}.
\newblock


\bibitem[Kotaru et~al\mbox{.}(2015)]%
        {kotaru2015spotfi}
\bibfield{author}{\bibinfo{person}{Manikanta Kotaru}, \bibinfo{person}{Kiran
  Joshi}, \bibinfo{person}{Dinesh Bharadia}, {and} \bibinfo{person}{Sachin
  Katti}.} \bibinfo{year}{2015}\natexlab{}.
\newblock \showarticletitle{Spotfi: Decimeter level localization using wifi}.
  In \bibinfo{booktitle}{\emph{SIGCOMM}}.
\newblock


\bibitem[Kwon et~al\mbox{.}(2020)]%
        {kwon2020imutube}
\bibfield{author}{\bibinfo{person}{Hyeokhyen Kwon}, \bibinfo{person}{Catherine
  Tong}, \bibinfo{person}{Harish Haresamudram}, \bibinfo{person}{Yan Gao},
  \bibinfo{person}{Gregory~D Abowd}, \bibinfo{person}{Nicholas~D Lane}, {and}
  \bibinfo{person}{Thomas Ploetz}.} \bibinfo{year}{2020}\natexlab{}.
\newblock \showarticletitle{Imutube: Automatic extraction of virtual on-body
  accelerometry from video for human activity recognition}.
\newblock \bibinfo{journal}{\emph{ACM Interact. Mob. Wearable Ubiquitous
  Technol.}} \bibinfo{volume}{4}, \bibinfo{number}{3} (\bibinfo{year}{2020}),
  \bibinfo{pages}{1--29}.
\newblock


\bibitem[Li and Tam(1998)]%
        {li1998iterative}
\bibfield{author}{\bibinfo{person}{CH Li} {and} \bibinfo{person}{Peter
  Kwong-Shun Tam}.} \bibinfo{year}{1998}\natexlab{}.
\newblock \showarticletitle{An iterative algorithm for minimum cross entropy
  thresholding}.
\newblock \bibinfo{journal}{\emph{Pattern recognition letters}}
  \bibinfo{volume}{19}, \bibinfo{number}{8} (\bibinfo{year}{1998}),
  \bibinfo{pages}{771--776}.
\newblock


\bibitem[Liu et~al\mbox{.}(2022)]%
        {liuvi}
\bibfield{author}{\bibinfo{person}{Hansi Liu}, \bibinfo{person}{Abrar Alali},
  \bibinfo{person}{Mohamed Ibrahim}, \bibinfo{person}{Bryan~Bo Cao},
  \bibinfo{person}{Nicholas Meegan}, \bibinfo{person}{Hongyu Li},
  \bibinfo{person}{Marco Gruteser}, \bibinfo{person}{Shubham Jain},
  \bibinfo{person}{Kristin Dana}, \bibinfo{person}{Ashwin Ashok},
  \bibinfo{person}{Bin Cheng}, {and} \bibinfo{person}{Hongsheng Lu}.}
  \bibinfo{year}{2022}\natexlab{}.
\newblock \showarticletitle{Vi-Fi: Associating Moving Subjects across Vision
  and Wireless Sensors}. In \bibinfo{booktitle}{\emph{IPSN}}.
\newblock


\bibitem[Liu et~al\mbox{.}(2021)]%
        {videoImu}
\bibfield{author}{\bibinfo{person}{Yilin Liu}, \bibinfo{person}{Shijia Zhang},
  {and} \bibinfo{person}{Mahanth Gowda}.} \bibinfo{year}{2021}\natexlab{}.
\newblock \showarticletitle{When Video meets Inertial Sensors: Zero-shot Domain
  Adaptation for Finger Motion Analytics with Inertial Sensors}. In
  \bibinfo{booktitle}{\emph{IoTDI}}.
\newblock


\bibitem[Livshitz(2017)]%
        {livshitz2017tracking}
\bibfield{author}{\bibinfo{person}{Michael Livshitz}.}
  \bibinfo{year}{2017}\natexlab{}.
\newblock \showarticletitle{Tracking radar targets with multiple reflection
  points}.
\newblock \bibinfo{journal}{\emph{Texas Instruments Application Note}}
  (\bibinfo{year}{2017}).
\newblock


\bibitem[Lu et~al\mbox{.}(2020)]%
        {lu2020milliego}
\bibfield{author}{\bibinfo{person}{Chris~Xiaoxuan Lu}, \bibinfo{person}{Muhamad
  Risqi~U Saputra}, \bibinfo{person}{Peijun Zhao}, \bibinfo{person}{Yasin
  Almalioglu}, \bibinfo{person}{Pedro~PB de Gusmao}, \bibinfo{person}{Changhao
  Chen}, \bibinfo{person}{Ke Sun}, \bibinfo{person}{Niki Trigoni}, {and}
  \bibinfo{person}{Andrew Markham}.} \bibinfo{year}{2020}\natexlab{}.
\newblock \showarticletitle{milliEgo: single-chip mmWave radar aided egomotion
  estimation via deep sensor fusion}. In \bibinfo{booktitle}{\emph{SenSys}}.
\newblock


\bibitem[Lu(2021)]%
        {fleeting-transmission}
\bibfield{author}{\bibinfo{person}{Donna Lu}.} \bibinfo{year}{2021}\natexlab{}.
\newblock \bibinfo{title}{Covid Delta variant is in the air you breathe}.
\newblock
\newblock
\newblock
\shownote{The Guardian}.


\bibitem[Madgwick et~al\mbox{.}(2011)]%
        {madgwick2011Imutrack}
\bibfield{author}{\bibinfo{person}{S.~OH Madgwick}, \bibinfo{person}{A.~JL
  Harrison}, {and} \bibinfo{person}{R. Vaidyanathan}.}
  \bibinfo{year}{2011}\natexlab{}.
\newblock \showarticletitle{Estimation of IMU and MARG orientation using a
  gradient descent algorithm}. In \bibinfo{booktitle}{\emph{ICORR}}.
\newblock


\bibitem[Mu et~al\mbox{.}(2020)]%
        {imuAdaption}
\bibfield{author}{\bibinfo{person}{Fangzhi Mu}, \bibinfo{person}{Xiao Gu},
  \bibinfo{person}{Yao Guo}, {and} \bibinfo{person}{Benny Lo}.}
  \bibinfo{year}{2020}\natexlab{}.
\newblock \showarticletitle{Unsupervised Domain Adaptation for
  Position-Independent IMU Based Gait Analysis}. In
  \bibinfo{booktitle}{\emph{IEEE Sensors J.}}
\newblock


\bibitem[Nguyen et~al\mbox{.}(2014)]%
        {identitylink}
\bibfield{author}{\bibinfo{person}{Le~T. Nguyen}, \bibinfo{person}{Yu~Seung
  Kim}, \bibinfo{person}{Patrick Tague}, {and} \bibinfo{person}{Joy Zhang}.}
  \bibinfo{year}{2014}\natexlab{}.
\newblock \showarticletitle{IdentityLink: User-Device Linking through Visual
  and RF-Signal Cues}. In \bibinfo{booktitle}{\emph{UbiComp}}.
\newblock


\bibitem[Pan et~al\mbox{.}(2018)]%
        {pan2018universense}
\bibfield{author}{\bibinfo{person}{Shijia Pan}, \bibinfo{person}{Carlos Ruiz},
  \bibinfo{person}{Jun Han}, \bibinfo{person}{Adeola Bannis},
  \bibinfo{person}{Patrick Tague}, \bibinfo{person}{Hae~Young Noh}, {and}
  \bibinfo{person}{Pei Zhang}.} \bibinfo{year}{2018}\natexlab{}.
\newblock \showarticletitle{Universense: Iot device pairing through
  heterogeneous sensing signals}. In \bibinfo{booktitle}{\emph{HotMobile}}.
\newblock


\bibitem[Park et~al\mbox{.}(2017)]%
        {park2017colored}
\bibfield{author}{\bibinfo{person}{Jaesik Park}, \bibinfo{person}{Qian-Yi
  Zhou}, {and} \bibinfo{person}{Vladlen Koltun}.}
  \bibinfo{year}{2017}\natexlab{}.
\newblock \showarticletitle{Colored point cloud registration revisited}. In
  \bibinfo{booktitle}{\emph{ICCV}}.
\newblock


\bibitem[Patrick~Rathje1(2022)]%
        {traceband}
\bibfield{author}{\bibinfo{person}{Olaf~Landsiedel1 Patrick~Rathje1}.}
  \bibinfo{year}{2022}\natexlab{}.
\newblock \showarticletitle{traceband:privacy preserving contact tracing on
  low-power wristband}. In \bibinfo{booktitle}{\emph{EWSN}}.
\newblock


\bibitem[Pegoraro et~al\mbox{.}(2020)]%
        {pegoraro2020multiperson}
\bibfield{author}{\bibinfo{person}{Jacopo Pegoraro}, \bibinfo{person}{Francesca
  Meneghello}, {and} \bibinfo{person}{Michele Rossi}.}
  \bibinfo{year}{2020}\natexlab{}.
\newblock \showarticletitle{Multiperson continuous tracking and identification
  from mm-wave micro-Doppler signatures}.
\newblock \bibinfo{journal}{\emph{GRSS}} \bibinfo{volume}{59},
  \bibinfo{number}{4} (\bibinfo{year}{2020}), \bibinfo{pages}{2994--3009}.
\newblock


\bibitem[Peng et~al\mbox{.}(2007)]%
        {peng2007beepbeep}
\bibfield{author}{\bibinfo{person}{C. Peng}, \bibinfo{person}{G. Shen},
  \bibinfo{person}{Y. Zhang}, \bibinfo{person}{Y. Li}, {and}
  \bibinfo{person}{K. Tan}.} \bibinfo{year}{2007}\natexlab{}.
\newblock \showarticletitle{Beepbeep: a high accuracy acoustic ranging system
  using cots mobile devices}. In \bibinfo{booktitle}{\emph{SenSys}}.
\newblock


\bibitem[Qi et~al\mbox{.}(2017)]%
        {pointnet}
\bibfield{author}{\bibinfo{person}{Charles~R Qi}, \bibinfo{person}{Hao Su},
  \bibinfo{person}{Kaichun Mo}, {and} \bibinfo{person}{Leonidas~J Guibas}.}
  \bibinfo{year}{2017}\natexlab{}.
\newblock \showarticletitle{Pointnet: Deep learning on point sets for 3d
  classification and segmentation}. In \bibinfo{booktitle}{\emph{CVPR}}.
\newblock


\bibitem[Redmon et~al\mbox{.}(2016)]%
        {yolo}
\bibfield{author}{\bibinfo{person}{Joseph Redmon}, \bibinfo{person}{Santosh
  Divvala}, \bibinfo{person}{Ross Girshick}, {and} \bibinfo{person}{Ali
  Farhadi}.} \bibinfo{year}{2016}\natexlab{}.
\newblock \showarticletitle{You only look once: Unified, real-time object
  detection}. In \bibinfo{booktitle}{\emph{CVPR}}.
\newblock


\bibitem[Rey et~al\mbox{.}(2019)]%
        {rey2019rgbV2Imu}
\bibfield{author}{\bibinfo{person}{Vitor~Fortes Rey}, \bibinfo{person}{Peter
  Hevesi}, \bibinfo{person}{Onorina Kovalenko}, {and} \bibinfo{person}{Paul
  Lukowicz}.} \bibinfo{year}{2019}\natexlab{}.
\newblock \showarticletitle{Let there be IMU data: generating training data for
  wearable, motion sensor based activity recognition from monocular RGB
  videos}. In \bibinfo{booktitle}{\emph{UbiComp}}.
\newblock


\bibitem[Ruiz et~al\mbox{.}(2020)]%
        {ruiz2020idiot}
\bibfield{author}{\bibinfo{person}{Carlos Ruiz}, \bibinfo{person}{Shijia Pan},
  \bibinfo{person}{Adeola Bannis}, \bibinfo{person}{Ming-Po Chang},
  \bibinfo{person}{Hae~Young Noh}, {and} \bibinfo{person}{Pei Zhang}.}
  \bibinfo{year}{2020}\natexlab{}.
\newblock \showarticletitle{IDIoT: Towards ubiquitous identification of iot
  devices through visual and inertial orientation matching during human
  activity}. In \bibinfo{booktitle}{\emph{IoTDI}}.
\newblock


\bibitem[Schroff et~al\mbox{.}(2015)]%
        {facenet}
\bibfield{author}{\bibinfo{person}{Florian Schroff}, \bibinfo{person}{Dmitry
  Kalenichenko}, {and} \bibinfo{person}{James Philbin}.}
  \bibinfo{year}{2015}\natexlab{}.
\newblock \showarticletitle{Facenet: A unified embedding for face recognition
  and clustering}. In \bibinfo{booktitle}{\emph{CVPR}}.
\newblock


\bibitem[Shuai et~al\mbox{.}(2021)]%
        {shuai2021millieye}
\bibfield{author}{\bibinfo{person}{Xian Shuai}, \bibinfo{person}{Yulin Shen},
  \bibinfo{person}{Yi Tang}, \bibinfo{person}{Shuyao Shi},
  \bibinfo{person}{Luping Ji}, {and} \bibinfo{person}{Guoliang Xing}.}
  \bibinfo{year}{2021}\natexlab{}.
\newblock \showarticletitle{millieye: A lightweight mmwave radar and camera
  fusion system for robust object detection}. In
  \bibinfo{booktitle}{\emph{IoTDI}}.
\newblock


\bibitem[Vaswani et~al\mbox{.}(2017)]%
        {vaswani2017attention}
\bibfield{author}{\bibinfo{person}{Ashish Vaswani}, \bibinfo{person}{Noam
  Shazeer}, \bibinfo{person}{Niki Parmar}, \bibinfo{person}{Jakob Uszkoreit},
  \bibinfo{person}{Llion Jones}, \bibinfo{person}{Aidan~N Gomez},
  \bibinfo{person}{{\L}ukasz Kaiser}, {and} \bibinfo{person}{Illia
  Polosukhin}.} \bibinfo{year}{2017}\natexlab{}.
\newblock \showarticletitle{Attention is all you need}. In
  \bibinfo{booktitle}{\emph{NIPS}}.
\newblock


\bibitem[Wang et~al\mbox{.}(2019)]%
        {wang2019wipin}
\bibfield{author}{\bibinfo{person}{Fei Wang}, \bibinfo{person}{Jinsong Han},
  \bibinfo{person}{Feng Lin}, {and} \bibinfo{person}{Kui Ren}.}
  \bibinfo{year}{2019}\natexlab{}.
\newblock \showarticletitle{Wipin: Operation-free passive person identification
  using wi-fi signals}. In \bibinfo{booktitle}{\emph{GLOBECOM}}.
\newblock


\bibitem[Wojke and Bewley(2018)]%
        {deepsort}
\bibfield{author}{\bibinfo{person}{Nicolai Wojke} {and} \bibinfo{person}{Alex
  Bewley}.} \bibinfo{year}{2018}\natexlab{}.
\newblock \showarticletitle{Deep Cosine Metric Learning for Person
  Re-identification}. In \bibinfo{booktitle}{\emph{WACV}}.
\newblock


\bibitem[Wu et~al\mbox{.}(2020)]%
        {wu2020mmtrack}
\bibfield{author}{\bibinfo{person}{Chenshu Wu}, \bibinfo{person}{Feng Zhang},
  \bibinfo{person}{Beibei Wang}, {and} \bibinfo{person}{KJ~Ray Liu}.}
  \bibinfo{year}{2020}\natexlab{}.
\newblock \showarticletitle{mmTrack: Passive multi-person localization using
  commodity millimeter wave radio}. In \bibinfo{booktitle}{\emph{INFOCOM}}.
\newblock


\bibitem[Xiao et~al\mbox{.}(2016)]%
        {xiao2016survey}
\bibfield{author}{\bibinfo{person}{Jiang Xiao}, \bibinfo{person}{Zimu Zhou},
  \bibinfo{person}{Youwen Yi}, {and} \bibinfo{person}{Lionel~M Ni}.}
  \bibinfo{year}{2016}\natexlab{}.
\newblock \showarticletitle{A survey on wireless indoor localization from the
  device perspective}.
\newblock \bibinfo{journal}{\emph{ACM Comput. Surv}} \bibinfo{volume}{49},
  \bibinfo{number}{2} (\bibinfo{year}{2016}).
\newblock


\bibitem[Xu et~al\mbox{.}(2020)]%
        {xu2020edge}
\bibfield{author}{\bibinfo{person}{Jingao Xu}, \bibinfo{person}{Hao Cao},
  \bibinfo{person}{Danyang Li}, \bibinfo{person}{Kehong Huang},
  \bibinfo{person}{Chen Qian}, \bibinfo{person}{Longfei Shangguan}, {and}
  \bibinfo{person}{Zheng Yang}.} \bibinfo{year}{2020}\natexlab{}.
\newblock \showarticletitle{Edge assisted mobile semantic visual slam}. In
  \bibinfo{booktitle}{\emph{INFOCOM}}.
\newblock


\bibitem[Yan et~al\mbox{.}(2018)]%
        {ridi}
\bibfield{author}{\bibinfo{person}{Hang Yan}, \bibinfo{person}{Qi Shan}, {and}
  \bibinfo{person}{Yasutaka Furukawa}.} \bibinfo{year}{2018}\natexlab{}.
\newblock \showarticletitle{RIDI: Robust IMU double integration}. In
  \bibinfo{booktitle}{\emph{ECCV}}.
\newblock


\bibitem[Yang et~al\mbox{.}(2020)]%
        {yang2020mu}
\bibfield{author}{\bibinfo{person}{Xin Yang}, \bibinfo{person}{Jian Liu},
  \bibinfo{person}{Yingying Chen}, \bibinfo{person}{Xiaonan Guo}, {and}
  \bibinfo{person}{Yucheng Xie}.} \bibinfo{year}{2020}\natexlab{}.
\newblock \showarticletitle{MU-ID: Multi-user identification through gaits
  using millimeter wave radios}. In \bibinfo{booktitle}{\emph{INFOCOM}}.
\newblock


\bibitem[Yun et~al\mbox{.}(2017)]%
        {yun2017strata}
\bibfield{author}{\bibinfo{person}{Sangki Yun}, \bibinfo{person}{Yi-Chao Chen},
  \bibinfo{person}{Huihuang Zheng}, \bibinfo{person}{Lili Qiu}, {and}
  \bibinfo{person}{Wenguang Mao}.} \bibinfo{year}{2017}\natexlab{}.
\newblock \showarticletitle{Strata: Fine-grained acoustic-based device-free
  tracking}. In \bibinfo{booktitle}{\emph{MobiCom}}.
\newblock


\bibitem[Zeng et~al\mbox{.}(2016)]%
        {zeng2016wiwho}
\bibfield{author}{\bibinfo{person}{Yunze Zeng}, \bibinfo{person}{Parth~H
  Pathak}, {and} \bibinfo{person}{Prasant Mohapatra}.}
  \bibinfo{year}{2016}\natexlab{}.
\newblock \showarticletitle{WiWho: WiFi-based person identification in smart
  spaces}. In \bibinfo{booktitle}{\emph{IPSN}}.
\newblock


\bibitem[Zhao et~al\mbox{.}(2019)]%
        {zhao2019mid}
\bibfield{author}{\bibinfo{person}{Peijun Zhao},
  \bibinfo{person}{Chris~Xiaoxuan Lu}, \bibinfo{person}{Jianan Wang},
  \bibinfo{person}{Changhao Chen}, \bibinfo{person}{Wei Wang},
  \bibinfo{person}{Niki Trigoni}, {and} \bibinfo{person}{Andrew Markham}.}
  \bibinfo{year}{2019}\natexlab{}.
\newblock \showarticletitle{mid: Tracking and identifying people with
  millimeter wave radar}. In \bibinfo{booktitle}{\emph{DCOSS}}.
\newblock


\end{thebibliography}

\end{document}